\documentclass[prb,twocolumn,showpacs,superscriptaddress]{revtex4-1} 
\usepackage{color}
\usepackage{graphicx} 
\usepackage{epstopdf}
\usepackage[colorlinks,linkcolor=blue,anchorcolor=blue,citecolor=blue,urlcolor=blue]{hyperref}

\newcommand{\dt}{\delta{t}}
\newcommand{\Tr}{\mathrm{Tr}}

\def\be{\begin{equation}}
\def\ee{\end{equation}}
\def\bea{\begin{eqnarray}}
\def\eea{\end{eqnarray}}

\def\nn{\nonumber}

\begin{document}
\title{Detecting edge degeneracy in interacting topological insulators 
through entanglement entropy}
\author{Da Wang}
\affiliation{Department of Physics, University of California, 
San Diego, California 92093, USA}
\affiliation{National Laboratory of Solid State Microstructures 
 $\&$ School of Physics, Nanjing University, Nanjing, 210093, China}
\author{Shenglong Xu}
\affiliation{Department of Physics, University of California, 
San Diego, California 92093, USA}
\author{Yu Wang}
\affiliation{School of Physics and Technology, Wuhan University, Wuhan 430072, China}
\author{Congjun Wu}
\affiliation{Department of Physics, University of California, 
San Diego, California 92093, USA}
\begin{abstract}
The existence of degenerate or gapless edge states is a characteristic 
feature of topological insulators, but is difficult to detect in
the presence of interactons.
We propose a new method to obtain the degeneracy of the edge states 
from the perspective of entanglement entropy, which is very useful
to identify interacting topological states.
Employing the determinant quantum Monte Carlo technique, we investigate 
the interaction effect on two representative models of 
fermionic topological insulators in one and two dimensions, respectively. 
In the two topologically nontrivial phases, the edge degeneracies 
are reduced by interactions but remain to be nontrivial. 
\end{abstract}
\pacs{03.65.Vf, 03.65.Ud, 02.70.Ss, 71.10.Fd}
\maketitle

\section{Introduction}
Topologically nontrivial states of matter are a central topic of
condensed matter physics, which are classified to two categories
according to their ground state entanglement properties.
The long-range entangled topological state, often named 
topologically ordered state, is characterized by the ground 
state degeneracy on a closed manifold \cite{Wen1990};
the short-range entangled topological insulator
can be characterized by its edge degeneracy on an open boundary. 
(Below we use the term {\it topological insulator} to 
represent the general short-range entangled topological states 
protected by certain symmetries \cite{Schnyder2008,*Gu2009},  
not just for the time-reversal invariant $\mathcal{Z}_2$ 
topological insulators \cite{Hasan2010,*Qi2011}.) 
For a non-interacting topological insulator, edge degeneracy 
comes directly from the zero energy edge mode, which is 
protected by its bulk topological property through the bulk-edge 
correspondence \cite{Hatsugai1993,*Ryu2002,Qi2006}
\footnote{In some special cases, there is no zero mode on an open 
boundary even in a topologically nontrivial state, e.g. 
Refs.~\cite {Turner2010,*Hughes2011,Huang2012}. 
Then the bulk-edge correspondence should be generalized by introducing 
twisted boundary condition \cite{Qi2006}.}.

However, the single-particle picture of the edge zero modes does not 
apply in interacting systems. 
The usual bulk-edge correspondence should be understood as the 
relation between bulk topological property and the many-body 
ground state degeneracy on the edge 
\footnote{There is a subtlety when we say the degeneracy of a 
''gapless'' system. We emphasize that it depends on the sequence 
of two limits: zero temperature and infinite lattice size 
\cite{Castelnovo2007}. In this article, we take zero temperature 
limit first.}.
The concept of edge degeneracy originates from the study 
of critical quantum systems \cite{Affleck1991}. 
Recently, it was also generalized to topologically ordered
systems \cite{Wang2012b}, and has 
been widely used for the classification 
of interacting topological insulators 
\cite{Fidkowski2011,*Qi2013,*Yao2013,Tang2012}. 
In this article, we will apply this concept to the systems of 
interacting topological insulators. 

Edge properties are important for the study of interaction effects in 
topological insulators. 
The first problem studied is the edge stability in the time-reversal invariant 
topological insulators in the presence of strong interactions 
\cite{Wu2006,*Xu2006}.
Because the edge is gapless, interaction effects on the edge 
are more prominent than those in the bulk, which can lead to edge 
instabilities while maintain the time-reversal invariance in the bulk.
The above picture has also been confirmed in quantum Monte Carlo 
(QMC) simulations \cite{Zheng2011,Hohenadler2011}. 
In recent years, interacting topological insulators have been 
intensively studied \cite{Wu2006,
*Xu2006,Hohenadler2011,Zheng2011,Yu2011,Raghu2008,*Dzero2010,
*Shitade2009,*Zhang2012,*Rachel2010,*Varney2010,*Yuan2012},
which have also been classified according to different 
symmetries \cite{Fidkowski2011,*Qi2013,
*Yao2013,Tang2012,Chen2011,*Turner2011,*Lu2012,*Gu2012,*Wang2013,Tang2012}. 
For the time-reversal invariant topological insulators, the $\mathcal{Z}_2$ index
can be formulated in terms of the single-particle Green's 
functions \cite{Wang2010,*Wang2011,*Wang2012}, which has 
been calculated based on both  analytic and numeric methods
\cite{Manmana2012,Wang2012a,*Araujo2013,Hung2013,*Lang2013}. 

On the other hand, quantum entanglement provides a particular 
perspective to investigate quantum many-body physics 
\cite{Amico2008,*Eisert2010}. 
Entanglement entropy  measures non-local correlations 
between part $A$ and the rest of the system denoted as part $B$.
Entanglement entropy can be defined as the von 
Neumann entropy 
\bea
S_v=-\Tr[\rho_A\ln\rho_A],
\eea
based on the reduced density matrix $\rho_A=\Tr_B(\rho_{A\cup B})$.
For systems characterized by short-range entanglement, entanglement 
entropy obeys an area law: It is proportional to the area/length of 
the boundary, i.e., the entanglement cut.
However, in the quantum critical region, entanglement entropy shows a 
logarithmic dependence on the subsystem size due to the divergence of 
coherence length \cite{Vidal2003,*Calabrese2004}.
In topologically ordered systems with long-range entanglements,
a negative sub-leading term appears termed as topological 
entanglement entropy  \cite{Kitaev2006,*Levin2006},
which depends on the degeneracy of ground states. 
Topological entanglement entropy has been used to identify different 
topological orders in quantum spin liquid systems
\cite{Zhang2011,*Zhang2012a,Yan2011,*Jiang2012,*Jiang2012a}. 
For short-range entangled topological insulators, 
the single-particle entanglement spectrum \cite{Li2008} is found to 
exhibit a ``zero mode''-like behavior \cite{Ryu2006,Fidkowski2010}
in the non-interacting case.
However, interaction invalids such a single-particle picture, and 
thus a more delicate method is required to describe
interacting topological insulators. 

In this article, we propose a method to determine the edge 
degeneracy using entanglement entropy.
A quantity termed edge entanglement entropy $S_{n,\rm edge}$ (defined 
in Eq.~\ref{eq:sedge}) is employed to measure edge degeneracy 
for both non-interacting and interacting topological insulators. 
This work is motivated by a recently developed algorithm using 
the fermionic determinant QMC to calculate 
the Renyi entanglement entropy \cite{Grover2013,*Assaad2013}
\bea
S_n=-\frac{1}{n-1}\ln\Tr [\rho_A^n].
\eea
We employ this algorithm to study edge degeneracy of fermionic 
interacting topological insulators by measuring $S_{2,\rm edge}$ in
both one and two dimensional systems. 
Our methodology will be first explained by using 
the Su-Schrieffer-Heeger-Hubbard (SSHH) model \cite{Su1979}.
In the topologically nontrivial phase, the Hubbard $U$ reduces 
$S_{2,\rm edge}$ from $2\ln2$ to $\ln2$, corresponding to reducing
edge degeneracy from $4$ to $2$ in the thermodynamic limit. 
In 2D, $S_{2,\rm edge}$ also contributes a sub-leading term to the
entanglement entropy area law, as we observed in the Kane-Mele-Hubbard 
(KMH) model \cite{Kane2005} for a cylindrical geometry. 
Moreover, $S_{2,\rm edge}$ shows even-odd dependence on the system
size along the entanglement cut in agreement with the helical 
liquid behavior on the edge.

\section{The SSHH model}
The 1D SSHH model is defined as
\begin{eqnarray}
H_{\rm SSHH}&=&-\sum_{i=1,\sigma}^{2L}\left[t+\dt(-1)^i\right]
c_{i\sigma}^\dag c_{i+1,\sigma} + H.c. \nonumber \\
&& +\sum_{i=1}^{2L} \frac{U}{2} ( n_i-1 )^2 ,
\end{eqnarray}
where $\sigma=\uparrow,\downarrow$;
$n_i=\sum_\sigma c_{i\sigma}^\dag c_{i\sigma}$,;
$\dt$ controls the hopping dimerization strength;
$t$ is set $1$ below; $U$ is the Hubbard interaction.

\begin{figure}
\includegraphics[width=0.8\linewidth]{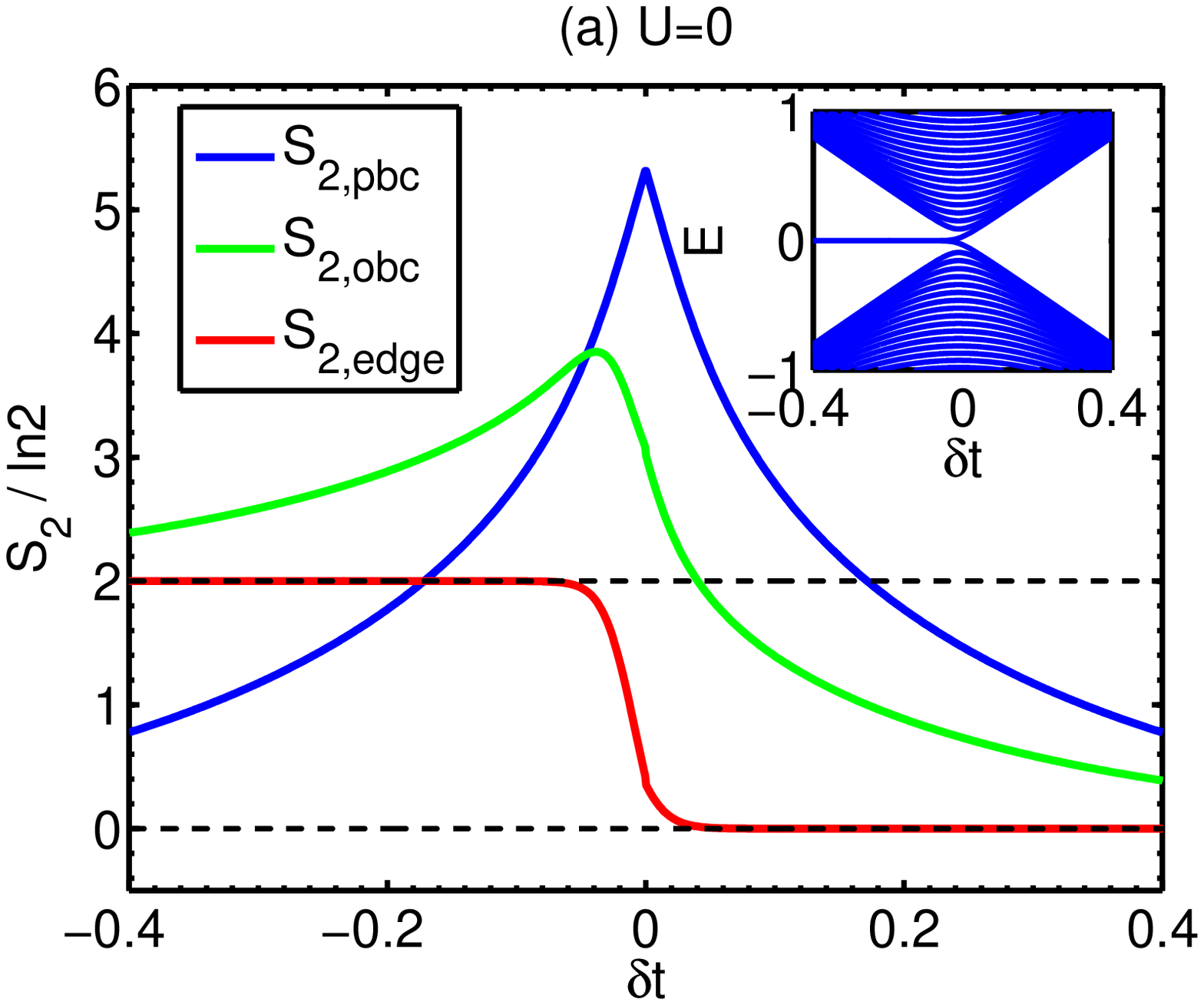}
\includegraphics[width=0.8\linewidth]{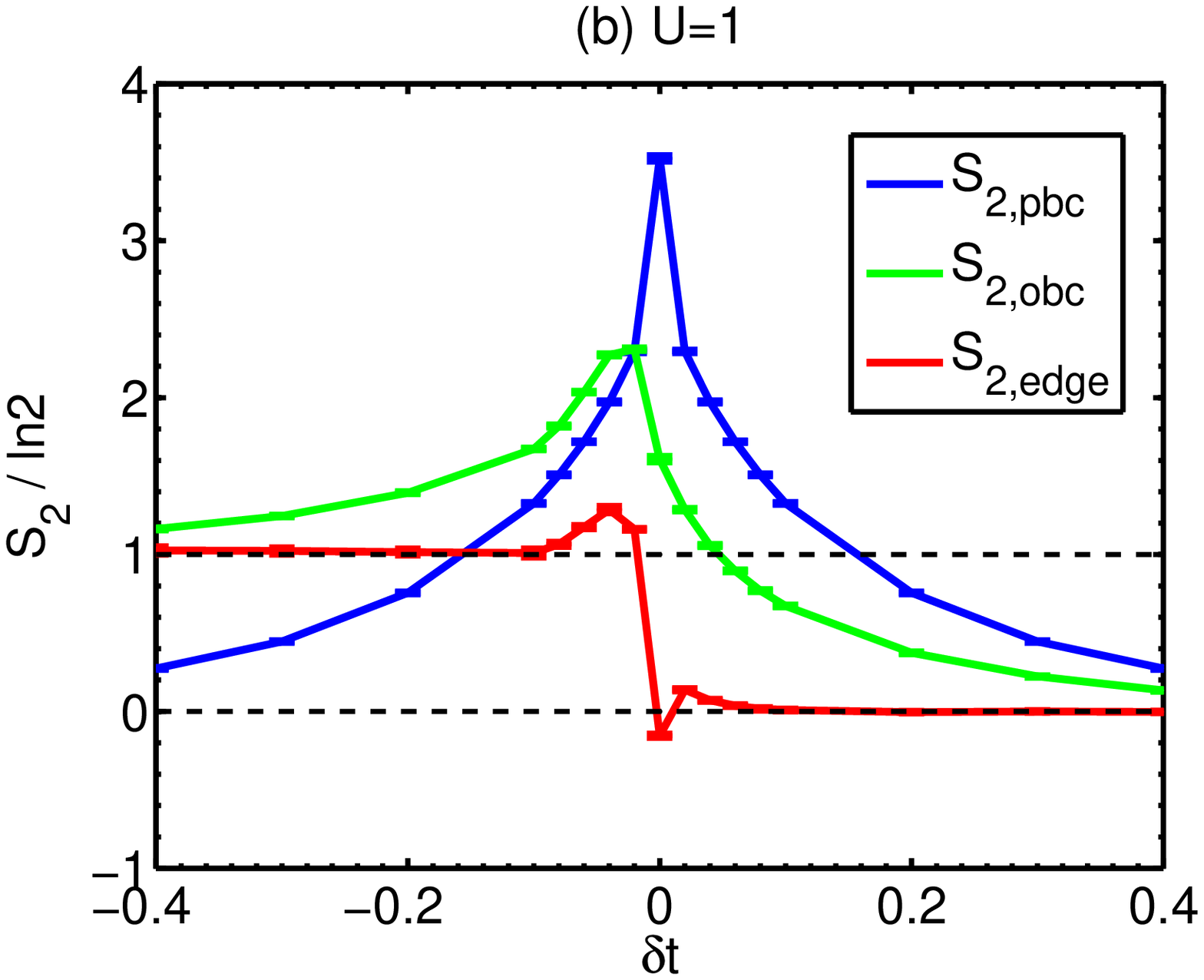}
\includegraphics[width=0.8\linewidth]{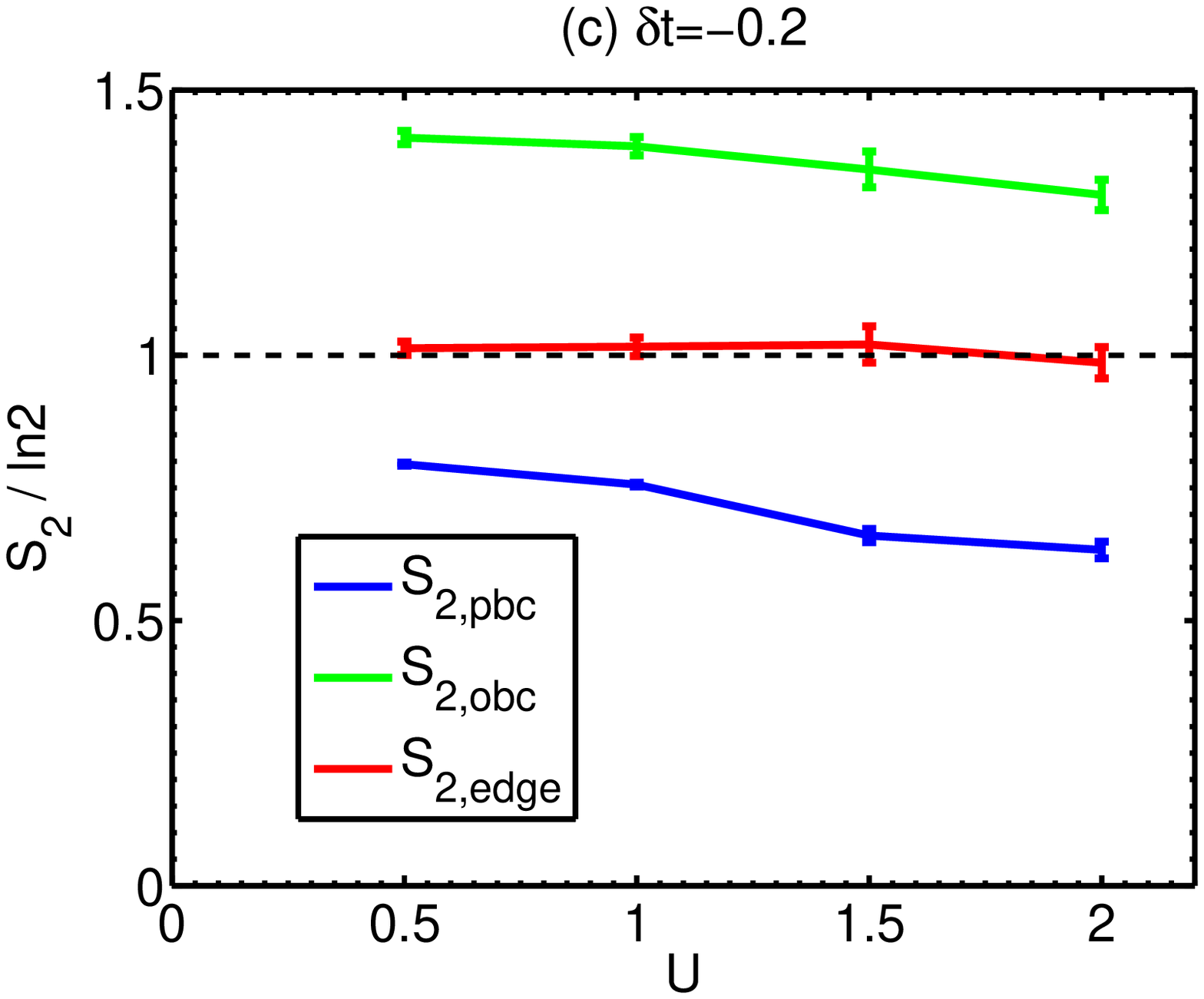}
\caption{
The 2nd order Renyi entanglement entropy of the 1D SSHH model.
$S_{2,\rm pbc}$, $S_{2,\rm obc}$, and $S_{2,\rm edge}$ are plotted in 
(a) (U=0), (b) (U=1) as functions of $\dt$, and in 
(c) ($\dt=-0.2$) as functions of $U$. 
Parameter values for the QMC simulations are the projection 
time $\beta=120$ ($200$) for $U\geq 1$ ($U=0.5$) and the discrete time step $\Delta\tau=0.1$.
The chain length is $L=100$ in (a), and $L=40$ in (b),(c). 
}
\label{fig:sshh}
\end{figure}

At $U=0$, this model is well-known exhibiting two topologically 
distinct ground states, characterized by Berry phase $0$ ($\dt>0$) 
and $\pi$ ($\dt<0$), respectively.
We use the convention that two sites $2i-1$ (odd) and $2i$ (even) are
combined into one unit cell.
The $\pi$-valued Berry phase guarantees the existence of 
one zero energy mode for each spin on each end  \cite{Ryu2006} 
(inset of Fig.~\ref{fig:sshh}(a)) at $\dt<0$.  
In order to study the entanglement entropy, the chain is
divided into two parts $A$ and $B$: 
Using the truncated correlation matrix $\mathcal{C}_{ij,\sigma}=
\langle c_{i\sigma}^\dag c_{j\sigma} \rangle$ with $i,j\in A$, 
the Renyi entanglement entropy can be obtained 
\cite{Peschel2003,*Cheong2004} as 
\bea
S_n=-\frac{1}{n-1}\sum\ln f_{i\sigma}^n,
\eea
where $f_{i\sigma}$ are eigenvalues of $\mathcal{C}_{ij,\sigma}$. 
$S_2$ is calculated under both the periodic boundary condition (PBC) and 
the open boundary condition (OBC), respectively, as 
plotted in Fig. ~\ref{fig:sshh}(a). 
In partice, for $\dt >0$, we choose $[1;L/2]$ as subsystem $A$
and $[L/2+1;L]$ as sublattice B for both OBC and PBC.
In this case, all the cuts are at weak bonds.
On the other hand, for $\dt <0$, we use a different partition
method such that the cuts are still at weak bonds.
For the case of OBC, we choose $[1;L/2+1]$ and $[L/2+2;L]$ as subsystems
A and B, respectively, while for the case of PBC, 
we choose $[2;L/2+1]$ and $[L/2+2;1]$ as subsystems A and B,
respectively, such that again all the cuts are at weak bonds.
Fig.~\ref{fig:sshh}(a) shows that neither PBC nor OBC gives quantized
entanglement entropy because of 
the short-range entanglement near the cut to bipartite
the system \cite{Ryu2006}. 
Of course, that we can also choose the cuts on strong bonds
and define the edge entanglement 

\begin{figure}
\includegraphics[width=0.9\linewidth]{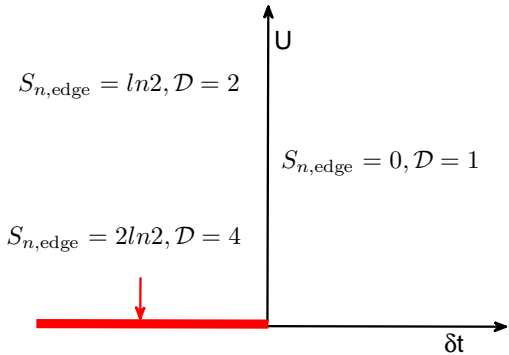}
\caption{The phase diagram of the SSHH model determined 
from the edge entanglement entropy $S_{n,\rm edge}$ and edge degeneracy $\mathcal{D}$.
}
\label{fig:sshhphase}
\end{figure}

To extract the entanglement between two edges, we define the
edge entanglement entropy as
\begin{eqnarray}
S_{n,\rm edge}=S_{n,\rm obc}-\frac{S_{n,\rm pbc}}{2} ,
\label{eq:sedge}
\end{eqnarray}
where half of $S_{n,\rm pbc}$ is subtracted because there are two cuts 
for defining entanglement in the case of PBC but only one in the case of OBC.
This definition also applies for the interacting case.
Eq.~\ref{eq:sedge} measures the nonlocal entanglement between the edges. 
Although $n$ can be any integer number, we only consider the
case of $n=2$ below because of the numerical convenience 
by QMC. 
Certainly we can also choose cuts on strong bonds and define
$S_{n,edge}$ by subtraction accordingly, the results of
the quantization of $S_{n,edge}$ remain robust.

The edge entanglement entropy exhibits a quantized behavior in 
two gapped phases.
At $U=0$, Fig. ~\ref{fig:sshh} (a) shows $S_{2,\rm edge}=2\ln2=\ln 4$ at 
$\dt<0$ while $S_{2,\rm edge}=0$ at $\dt>0$.
This result can be understood as follows.
For each spin component $\sigma$, two zero modes $\gamma_{(L,R)\sigma}$ 
at two ends are coupled through an effective hopping 
$t_{eff}\sim \exp(-\dt L)$, and then 
the bonding state, 
$\frac{1}{\sqrt 2}
(\gamma_{L\sigma}^\dag+\gamma_{R\sigma}^\dag)|0\rangle$,
contributes a $\ln2$ to $S_{n,\rm edge}$ in each spin component. 
More explicitly, this bonding edge states are occupied by both
spin components, i.e., the two-particle edge states can be 
written as
\bea
\frac{1}{2}
(\gamma_{L\uparrow}^\dag\gamma^\dag_{L\downarrow}
+\gamma_{R\uparrow}^\dag\gamma^\dagger_{R\downarrow}
+\gamma_{L\uparrow}^\dag\gamma^\dag_{R\downarrow}
+\gamma_{R\uparrow}^\dag\gamma^\dagger_{L\downarrow}
)|0\rangle,
\label{eq:entangleU0}
\eea
which clearly exhibit the $\ln4$ contribution to the edge entanglement
entropy.

Now let us consder to take the limit of $L\rightarrow +\infty$ in which
$t_{eff}$ approaches to zero and the edge modes become 
exactly zero modes.
Then each term in Eq. \ref{eq:entangleU0} corresponds to a zero 
mode for either edge. 
Say, after tracing out the degree of freedom on the right edge,
we arrive at the zero modes at the left edge as
$\gamma^\dag_{L\uparrow}\gamma^\dag_{L\downarrow}|0\rangle$,
$|0\rangle$, $\gamma^\dag_{L\uparrow}|0\rangle$,
$\gamma^\dag_{L\downarrow}|0\rangle$.
Thus the above defined $S_{2,edge}$ can be used as a topological index,
which corresponds to the thermodynamic entropy at zero temperature
of one edge. 
This explains the relation between entanglement entropy and the ground 
state degeneracy $\mathcal{D}$ on one open end as
\begin{eqnarray}
\ln \mathcal{D}=\lim_{L\to\infty}S_{n,\rm edge}(L),
\end{eqnarray} 
which converges to the same value independent of $n$. 
It is sufficient to calculate $S_2$ to determine edge degeneracy. 
A similar quantity to Eq.~\ref{eq:sedge} was used
as a topological invariant to study 1D non-interacting 
$p$-wave superconductor \cite{Kim2013a}. 
The physical meaning of this topological invariant becomes clear 
in our approach: it represents entanglement between two edges 
for a finite lattice size, and converges to edge degeneracy 
in the thermodynamic limit very quickly.
Note that at the critical point $\dt=0$, 
$S_{2,\rm edge}$ is negative and unquantized, in agreement with the 
``non-integer'' edge degeneracy in critical quantum systems \cite{Affleck1991}.

Now let us turn on the Hubbard interaction $U$. 
We combine the zero temperature projector QMC \cite{Assaad2008} 
with the new developed algorithm to calculate $S_2$ \cite{Grover2013} 
under PBC and OBC, respectively.
Due to the particle-hole symmetry, the half-filled SSHH model is 
free of the sign problem, and thus the QMC simulation can be
performed in a controllable way.
The results of
$S_2$ v.s. $\dt$ at $L=40$ and $U=1$ are calculated and 
plotted in Fig.~\ref{fig:sshh} (b). 
The behavior of $S_{2,\rm edge}$ is similar to the case of $U=0$ 
in Fig.~\ref{fig:sshh} (a)
except that its quantized value becomes $\ln2$ when $\dt<0$.
At large values of $L$, $U\gg t_{eff}$, and thus the singlet ground 
state changes to 
\bea
\frac{1}{\sqrt{2}}\left[\gamma_{L\uparrow}^\dag\gamma_{R\downarrow}^\dag
-\gamma_{L\downarrow}^\dag\gamma_{R\uparrow}^\dag\right]|0\rangle,
\eea
leads to $S_{2,\rm edge}=\ln2$.
Again in the limit of $L\rightarrow +\infty$, the edge modes
become exactly zero modes. 
If we trace out the right edge, the zero modes left at the
left edge is $\gamma^\dag_\uparrow |0\rangle$,
and $\gamma^\dag_\downarrow |0\rangle$, which
means that edge degeneracy $D$ is reduced from $4$ to $2$ 
by the Hubbard $U$, i.e,  the double and empty occupations of the 
edge states are projected out. 
Due to the exponential decay of $t_{eff}$, the finite-size
effect of $S_{2,\rm edge}$ is weak.
It converges to $\ln \mathcal{D}$ quickly even before 
$L$ goes large.


We have also calculated $S_{2,\rm pbc}$, $S_{2,\rm obc}$ and $S_{2,\rm edge}$ 
at different values of $U$ ranging from $0.5$ to $2$ as shown in 
Fig.~\ref{fig:sshh} (c). 
$S_{2,\rm pbc}$ and $S_{2,\rm obc}$ are non-quantized, which decreases
as increasing $U$ due to the suppression of charge fluctuations
across the cuts.
Nevertheless, $S_{2,\rm edge}$ is pinned at $\ln 2$ regardless of 
different values of $U$ due to the exponential suppression of $t_{eff}$.  
In Fig.~\ref{fig:sshhphase}, we set up the phase diagram of the 
SSHH model using $S_{n,\rm edge}$ and edge degeneracy. 
Similar phase diagram has been obtained by calculating the bulk 
topological number $\mathcal{Z}=0 (2)$ for $\dt>0 (<0)$ using 
Green's functions extracted from the density matrix renormalization 
group \cite{Manmana2012}. 
Our study here further indicates the edge behavior: in the 
topologically nontrivial region, edge degeneracy is reduced 
from $4$ to $2$ by the Hubbard interaction \cite{Tang2012}
at half-filling.

When $U$ is large, the low energy physics of the
SSHH model is described by the spin-$1/2$ 
Heisenberg-Peierls model $H=\sum_{i} J_i\vec{S}_i
\cdot\vec{S}_{i+1}$, where 
$J_i=J$ or $J'$ for the odd or even bond, respectively \cite{Wang2013b}. 
Our study shows that the cases of $J'<J$ and $J'>J$ belong 
to topologically distinct phases. 
At $J^\prime>J$, there is a free local moment at one end resulting 
in a double edge degeneracy. 
The transition occurs at $J=J'$ consistent with the 
critical behavior of the spin $1/2$ Heisenberg model 
\cite{Cloizeaux1962,Haldane1983}.

For the above results, the particle-hole symmetry gives rise to zero energy 
edges in non-interacting cases. The finite size effect couples the two edge
states together and contributes to $S_{n,\rm edge}$. Even in the interacting 
case, our numeric simulations show that $S_{n,\rm edge}$ remains robust. When 
the particle-hole symmetry is gone, for the 1D case, both edge states are
not at zero energy. They are either both occupied or empty, and thus 
$S_{n,\rm edge}$ will be reduced to zero. Nevertheless, our
method still applies to the 2D Kane-Mele-Hubbard model because the
chemical potential crosses the 1D band of edge states. The single particle
states right at the chemical potential play the role of zero energy
states.

\section{The KMH model}
Next we move to 2D and investigate the KMH model on a honeycomb lattice
defined as
\begin{eqnarray}
H_{\rm KMH}&=&-\sum_{\langle i,j \rangle,\sigma}tc_{i\sigma}^\dag c_{j\sigma} + 
\sum_{\langle\langle i,j \rangle\rangle,\sigma}
i\lambda c_{i\sigma}^\dag \sigma c_{j\sigma} \nonumber\\
&& +\sum_{i} \frac{U}{2} ( n_i-1 )^2 ,
\end{eqnarray}
where $\lambda$ is the next-nearest neighbor spin-orbit coupling;
$\sigma=\uparrow,\downarrow$; again $t$ is set $1$. 
This model is free of the sign problem and has been investigated by 
the determinant QMC \cite{Zheng2011,Hohenadler2011}. 
Along the $y$-direction (zigzag), the PBC is applied, and 
along the $x$-direction (armchair), both of the  PBC and OBC are applied.
The PBC and OBC correspond to the toric and cylindrical geometries,
respectively. 
The lattice is divided into the subsystem $A$ with $1 \leq x \leq L/2$ 
and the environment $B$ with $L/2+1 \leq x \leq L$ for
the study of entanglement entropy.

\begin{figure}
\includegraphics[width=0.45\textwidth]{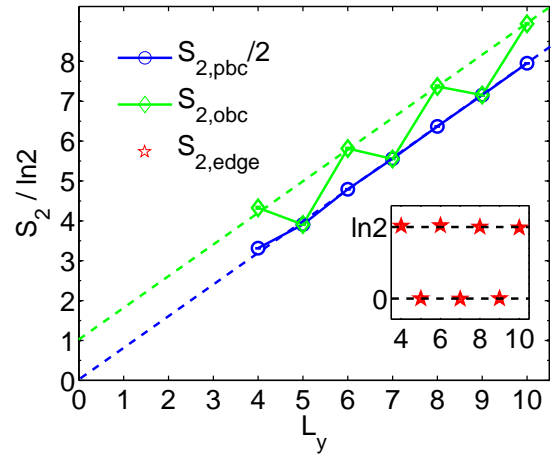}
\caption{The 2nd order Renyi entanglement entropy of the KMH model v.s. $L_y$. 
$S_{2,\rm edge}$ is plotted in the inset.
Parameter values are $U=1$, $\lambda=0.2$, $\beta=10L_y$, $\Delta\tau=0.1$, 
and $L_x=6$. }
\label{fig:kmhqmc}
\end{figure}

The QMC results of $S_{2,\rm pbc}$ and $S_{2,\rm obc}$ for the KMH model 
are shown in Fig.~\ref{fig:kmhqmc}.
$S_{2,\rm pbc}$ exhibits a standard area law, {\it i.e.}, $S_{2,\rm pbc}
\propto L_y$, while $S_{2,\rm obc}$ shows an even-odd oscillating behavior.
Then $S_{2,\rm edge}=\ln 2$ and $0$ for even and odd values of $L_y$, 
respectively, as shown in the inset. 
On the other hand, $S_{2,\rm edge}$ can also be obtained by extrapolating
$S_{2,\rm obc} (L_y)$ for even values of $L_y$, in which 
$S_{2,\rm edge}$ appears as the sub-leading term of the area law as
\bea
S_{2,\rm obc} (L_y)\approx \alpha L_y+S_{2,\rm edge} .
\eea
Such a sub-leading term is an analogy to the topological entanglement entropy in the long-range entangled
topological orders \cite{Kitaev2006,*Levin2006,Zhang2011,
*Zhang2012a,Yan2011,*Jiang2012,*Jiang2012a}.
We propose to use $S_{2,\rm edge}$ to characterize the short-range entangled 
topological insulators in 2D. 
Both topological entanglement entropy and $S_{2,\rm edge}$ are related to the ground state degeneracy, 
but account for bulk and edge states, respectively. 

\begin{figure}
\includegraphics[width=0.4\textwidth]{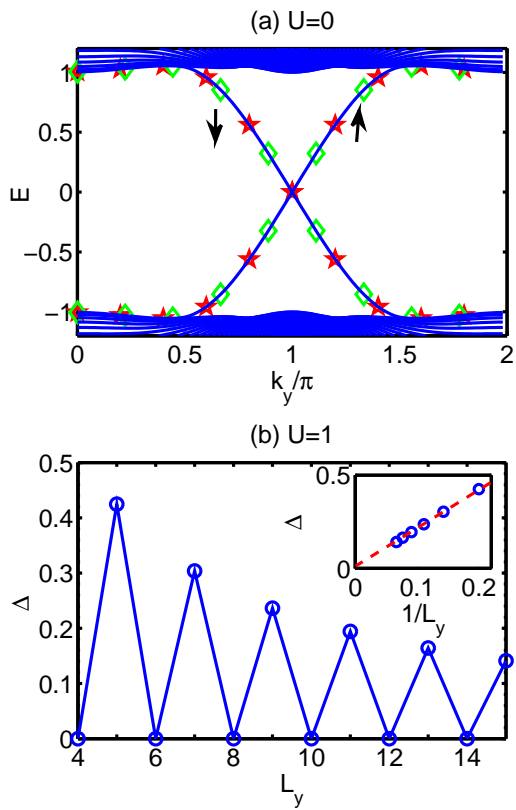}
\caption{The helical edge behavior of the KMH model. 
(a) The single-particle energy spectra at $U=0$. 
Red and blue symbols mark the edge specta with even ($L_y=10$)
and odd ($L_y=9$) values of $L_y$, respectively. 
(b) The many-body energy gap $\Delta$ v.s. $L_y$ for the 
effective model Eq.~\ref{eq:helical} with $U=1$ and 
$\Lambda=\pi v_F/2=1$. 
The inset plots $\Delta$ v.s. $1/L_y$ for only odd $L_y$.
}
\label{fig:kmhhelical}
\end{figure}

Next we explain the origin of the nonzero $S_{2,\rm edge}$ by analyzing the 
edge degeneracy. 
At $U=0$, such a behavior is a direct consequence of the zero-energy edge 
states, which has also been found in the Kitaev model \cite{Yao2010}, and 
non-interacting triplet topological superconductors \cite{Oliveira2013}. 
In Fig.~\ref{fig:kmhhelical} (a), the energy spectrum with the open edges
is plotted as a function of $k_y$ which is conserved due to the PBC
along the $y$-direction. 
The edge zero mode is located at $k_y=\pi$, which
is accessible for even values of $L_y$ but not for odd $L_y$, 
thus the many-body ground state degeneracy varies between 
$4$ and $1$ respectively. 

At $U>0$, the above single-particle picture does not hold any more. 
Interaction effects have to be fully taken into account to investigate 
the many-body edge degeneracy. 
We use an effective edge helical liquid defined in momentum 
space \cite{Wu2006},
\begin{eqnarray}
H_{hl}&=&\sum_{k,\sigma} \sigma v_F (k-\pi)c_{k\sigma}^\dag c_{k\sigma} 
\nn \\
&+& \frac{U}{L_y}\sum_{kk'q}c_{k+q\uparrow}^\dag c_{k\uparrow}
c_{k'-q\downarrow}^\dag c_{k'\downarrow} ,
\label{eq:helical}
\end{eqnarray}
where $c_{k\sigma}^\dag$ are creation operators of 
non-interacting edge states.
The $y$-direction momenta $k=\frac{2\pi n}{L_y}$ 
($n=0,1,\cdots,L_y-1$) are chosen only for edge states and 
satisfy $|v_F(k-\pi)|\leq\Lambda$ where $\Lambda$ is the energy cutoff. 
Since edge modes with different values of $k$ have different
localization lengths, rigorously speaking, the interaction 
matrix elements for the edge modes are $k$-dependent even for
the case of Hubbard model. 
Nevertheless, for simplicity, we neglect this dependence. 
In real calculations, we choose $\Lambda=\pi v_F/2=1$ without loss 
of generality. 
The number of momentum points for edge states within the cut off
is denoted as $N_k$.

Exact diagonalization method is employed to numerically solve
the many-body edge energy levels at $L_y\leq15$ ($N_k\leq8$), and the
energy gaps are plotted in Fig.~\ref{fig:kmhhelical} (b). 
For even $L_y$'s, $\Delta=0$ corresponds to a double degeneracy 
in agreement with the QMC result $S_{2,\rm edge}=\ln2$; 
For odd $L_y$'s, the ground state has no degeneracy, thus $S_{2,\rm edge}=0$. 
Nevertheless, the gap decreases to zero as increasing $L_y$ 
as shown in the inset. 
This gapless behavior in the thermodynamic limit has been obtained from 
the bosonization analysis \cite{Wu2006}, which shows that forward 
scattering does not open a gap in a helical liquid in the weak
interacting regime. 


\section{Summary}
We propose a quantized quantity of the edge entanglement entropy $S_{n,\rm edge}$ to determine 
edge degeneracy in topological insulators in the presence of interactions.
Using the fermionic quantum Monte Carlo algorithm, $S_{n,\rm edge}$ is 
calculated for both the interacting 1D SSHH model and 2D KMH model. 
In topologically nontrivial phases of these models, the Hubbard 
$U$ suppresses the quantized values of $S_{2,\rm edge}$ from $2\ln 2$
in the non-interacting cases to its half value $\ln 2$. 
In 2D, such a nonzero $S_{n,\rm edge}$ also contributes a sub-leading 
term in the entanglement entropy area law for a cylindrical geometry. 

Before closing this paper, some remarks are in order:
(I) Our QMC calculations are only performed at small and medium $U$. 
When $U$ goes large, the QMC numeric error of entanglement entropy increases 
significantly \cite{Assaad2013}.
Significant numeric efforts are needed to obtain reliable entanglement entropy. 
Very recently, a new QMC algorithm using the replica technique was 
proposed for fermionic systems \cite{Broecker2014}, which is 
more stable in the large $U$ regime and can be helpful to 
study the Mott transition regime in the future;
(II) If the third nearest neighbor hopping is added to the KMH model, 
two Dirac nodes are produced at $k_y=0$ and $\pi$ respectively 
on an edge \cite{Hung2013}. 
In this case, any small $U$ will gap out the edge states due 
to the Umklapp scattering \cite{Wu2006}.
Therefore, $S_{2,\rm edge}=0$ is expected to be consistent with the 
physical implication of the $\mathcal{Z}_2$ topological insulator.
(III) We have seen that the above edge state entanglement is built up 
through the effective coupling $\sim\exp(-L_x/\xi)$ between two edges,
in which $L_x$ is the width of the system, and $\xi$ is the typical
localization length of the edge modes.
On the other hand, for the 2D case, the length $L_y$ along the edge 
direction also gives another energy scale for the low energy edge
excitations, which is $\sim 1/L_y$. 
When $\exp(-L_x/\xi) \ll 1/L_y$ (the regime we are interest), for
the edge modes not right located at the Fermi energy, we can neglect
their contributions to the total EE, and only need to consider
the zero energy mode right at the Fermi energy. 
In this regime, our method applies.

\acknowledgements
D. W. thanks Zhoushen Huang and Xiao Chen for helpful discussions. 
This work is supported by NSF DMR-1410375 and AFOSR FA9550-14-1-0168.
Y.W. and C.W. acknowledges the financial support from the National 
Natural Science Foundation of China under Grant No. 11328403 and 
the Fundamental Research Funds for the Central Universities.
C. W. also acknowledges the support from the President’s 
Research Catalyst Awards of University of California. 
Part of the computational resources required for this work 
were accessed via the GlideinWMS \citep{Sfiligoi2009} 
on the Open Science Grid \citep{Pordes2007}.

\bibliography{entanglement}

\begin{thebibliography}{75}%
\makeatletter
\providecommand \@ifxundefined [1]{%
 \@ifx{#1\undefined}
}%
\providecommand \@ifnum [1]{%
 \ifnum #1\expandafter \@firstoftwo
 \else \expandafter \@secondoftwo
 \fi
}%
\providecommand \@ifx [1]{%
 \ifx #1\expandafter \@firstoftwo
 \else \expandafter \@secondoftwo
 \fi
}%
\providecommand \natexlab [1]{#1}%
\providecommand \enquote  [1]{``#1''}%
\providecommand \bibnamefont  [1]{#1}%
\providecommand \bibfnamefont [1]{#1}%
\providecommand \citenamefont [1]{#1}%
\providecommand \href@noop [0]{\@secondoftwo}%
\providecommand \href [0]{\begingroup \@sanitize@url \@href}%
\providecommand \@href[1]{\@@startlink{#1}\@@href}%
\providecommand \@@href[1]{\endgroup#1\@@endlink}%
\providecommand \@sanitize@url [0]{\catcode `\\12\catcode `\$12\catcode
  `\&12\catcode `\#12\catcode `\^12\catcode `\_12\catcode `\%12\relax}%
\providecommand \@@startlink[1]{}%
\providecommand \@@endlink[0]{}%
\providecommand \url  [0]{\begingroup\@sanitize@url \@url }%
\providecommand \@url [1]{\endgroup\@href {#1}{\urlprefix }}%
\providecommand \urlprefix  [0]{URL }%
\providecommand \Eprint [0]{\href }%
\providecommand \doibase [0]{http://dx.doi.org/}%
\providecommand \selectlanguage [0]{\@gobble}%
\providecommand \bibinfo  [0]{\@secondoftwo}%
\providecommand \bibfield  [0]{\@secondoftwo}%
\providecommand \translation [1]{[#1]}%
\providecommand \BibitemOpen [0]{}%
\providecommand \bibitemStop [0]{}%
\providecommand \bibitemNoStop [0]{.\EOS\space}%
\providecommand \EOS [0]{\spacefactor3000\relax}%
\providecommand \BibitemShut  [1]{\csname bibitem#1\endcsname}%
\let\auto@bib@innerbib\@empty
\bibitem [{\citenamefont {Wen}\ and\ \citenamefont {Niu}(1990)}]{Wen1990}%
  \BibitemOpen
  \bibfield  {author} {\bibinfo {author} {\bibfnamefont {X.~G.}\ \bibnamefont
  {Wen}}\ and\ \bibinfo {author} {\bibfnamefont {Q.}~\bibnamefont {Niu}},\
  }\href {\doibase 10.1103/PhysRevB.41.9377} {\bibfield  {journal} {\bibinfo
  {journal} {Phys. Rev. B}\ }\textbf {\bibinfo {volume} {41}},\ \bibinfo
  {pages} {9377} (\bibinfo {year} {1990})}\BibitemShut {NoStop}%
\bibitem [{\citenamefont {Schnyder}\ \emph {et~al.}(2008)\citenamefont
  {Schnyder}, \citenamefont {Ryu}, \citenamefont {Furusaki},\ and\
  \citenamefont {Ludwig}}]{Schnyder2008}%
  \BibitemOpen
  \bibfield  {author} {\bibinfo {author} {\bibfnamefont {A.~P.}\ \bibnamefont
  {Schnyder}}, \bibinfo {author} {\bibfnamefont {S.}~\bibnamefont {Ryu}},
  \bibinfo {author} {\bibfnamefont {A.}~\bibnamefont {Furusaki}}, \ and\
  \bibinfo {author} {\bibfnamefont {A.~W.~W.}\ \bibnamefont {Ludwig}},\ }\href
  {\doibase 10.1103/PhysRevB.78.195125} {\bibfield  {journal} {\bibinfo
  {journal} {Phys. Rev. B}\ }\textbf {\bibinfo {volume} {78}},\ \bibinfo
  {pages} {195125} (\bibinfo {year} {2008})}\BibitemShut {NoStop}%
\bibitem [{\citenamefont {Gu}\ and\ \citenamefont {Wen}(2009)}]{Gu2009}%
  \BibitemOpen
  \bibfield  {author} {\bibinfo {author} {\bibfnamefont {Z.-C.}\ \bibnamefont
  {Gu}}\ and\ \bibinfo {author} {\bibfnamefont {X.-G.}\ \bibnamefont {Wen}},\
  }\href {\doibase 10.1103/PhysRevB.80.155131} {\bibfield  {journal} {\bibinfo
  {journal} {Phys. Rev. B}\ }\textbf {\bibinfo {volume} {80}},\ \bibinfo
  {pages} {155131} (\bibinfo {year} {2009})}\BibitemShut {NoStop}%
\bibitem [{\citenamefont {Hasan}\ and\ \citenamefont {Kane}(2010)}]{Hasan2010}%
  \BibitemOpen
  \bibfield  {author} {\bibinfo {author} {\bibfnamefont {M.~Z.}\ \bibnamefont
  {Hasan}}\ and\ \bibinfo {author} {\bibfnamefont {C.~L.}\ \bibnamefont
  {Kane}},\ }\href {\doibase 10.1103/RevModPhys.82.3045} {\bibfield  {journal}
  {\bibinfo  {journal} {Rev. Mod. Phys.}\ }\textbf {\bibinfo {volume} {82}},\
  \bibinfo {pages} {3045} (\bibinfo {year} {2010})}\BibitemShut {NoStop}%
\bibitem [{\citenamefont {Qi}\ and\ \citenamefont {Zhang}(2011)}]{Qi2011}%
  \BibitemOpen
  \bibfield  {author} {\bibinfo {author} {\bibfnamefont {X.-L.}\ \bibnamefont
  {Qi}}\ and\ \bibinfo {author} {\bibfnamefont {S.-C.}\ \bibnamefont {Zhang}},\
  }\href {\doibase 10.1103/RevModPhys.83.1057} {\bibfield  {journal} {\bibinfo
  {journal} {Rev. Mod. Phys.}\ }\textbf {\bibinfo {volume} {83}},\ \bibinfo
  {pages} {1057} (\bibinfo {year} {2011})}\BibitemShut {NoStop}%
\bibitem [{\citenamefont {Hatsugai}(1993)}]{Hatsugai1993}%
  \BibitemOpen
  \bibfield  {author} {\bibinfo {author} {\bibfnamefont {Y.}~\bibnamefont
  {Hatsugai}},\ }\href {\doibase 10.1103/PhysRevLett.71.3697} {\bibfield
  {journal} {\bibinfo  {journal} {Phys. Rev. Lett.}\ }\textbf {\bibinfo
  {volume} {71}},\ \bibinfo {pages} {3697} (\bibinfo {year}
  {1993})}\BibitemShut {NoStop}%
\bibitem [{\citenamefont {Ryu}\ and\ \citenamefont {Hatsugai}(2002)}]{Ryu2002}%
  \BibitemOpen
  \bibfield  {author} {\bibinfo {author} {\bibfnamefont {S.}~\bibnamefont
  {Ryu}}\ and\ \bibinfo {author} {\bibfnamefont {Y.}~\bibnamefont {Hatsugai}},\
  }\href {\doibase 10.1103/PhysRevLett.89.077002} {\bibfield  {journal}
  {\bibinfo  {journal} {Phys. Rev. Lett.}\ }\textbf {\bibinfo {volume} {89}},\
  \bibinfo {pages} {077002} (\bibinfo {year} {2002})}\BibitemShut {NoStop}%
\bibitem [{\citenamefont {Qi}\ \emph {et~al.}(2006)\citenamefont {Qi},
  \citenamefont {Wu},\ and\ \citenamefont {Zhang}}]{Qi2006}%
  \BibitemOpen
  \bibfield  {author} {\bibinfo {author} {\bibfnamefont {X.-L.}\ \bibnamefont
  {Qi}}, \bibinfo {author} {\bibfnamefont {Y.-S.}\ \bibnamefont {Wu}}, \ and\
  \bibinfo {author} {\bibfnamefont {S.-C.}\ \bibnamefont {Zhang}},\ }\href
  {\doibase 10.1103/PhysRevB.74.045125} {\bibfield  {journal} {\bibinfo
  {journal} {Phys. Rev. B}\ }\textbf {\bibinfo {volume} {74}},\ \bibinfo
  {pages} {045125} (\bibinfo {year} {2006})}\BibitemShut {NoStop}%
\bibitem [{Note1()}]{Note1}%
  \BibitemOpen
  \bibinfo {note} {In some special cases, there is no zero mode on an open
  boundary even in a topologically nontrivial state, e.g. Refs.~\cite
  {Turner2010,*Hughes2011,Huang2012}. Then the bulk-edge correspondence should
  be generalized by introducing twisted boundary condition \cite
  {Qi2006}.}\BibitemShut {Stop}%
\bibitem [{Note2()}]{Note2}%
  \BibitemOpen
  \bibinfo {note} {There is a subtlety when we say the degeneracy of a
  ''gapless'' system. We emphasize that it depends on the sequence of two
  limits: zero temperature and infinite lattice size \cite {Castelnovo2007}. In
  this article, we take zero temperature limit first.}\BibitemShut {Stop}%
\bibitem [{\citenamefont {Affleck}\ and\ \citenamefont
  {Ludwig}(1991)}]{Affleck1991}%
  \BibitemOpen
  \bibfield  {author} {\bibinfo {author} {\bibfnamefont {I.}~\bibnamefont
  {Affleck}}\ and\ \bibinfo {author} {\bibfnamefont {A.~W.~W.}\ \bibnamefont
  {Ludwig}},\ }\href {\doibase 10.1103/PhysRevLett.67.161} {\bibfield
  {journal} {\bibinfo  {journal} {Phys. Rev. Lett.}\ }\textbf {\bibinfo
  {volume} {67}},\ \bibinfo {pages} {161} (\bibinfo {year} {1991})}\BibitemShut
  {NoStop}%
\bibitem [{\citenamefont {{Wang}}\ and\ \citenamefont
  {{Wen}}(2012)}]{Wang2012b}%
  \BibitemOpen
  \bibfield  {author} {\bibinfo {author} {\bibfnamefont {J.}~\bibnamefont
  {{Wang}}}\ and\ \bibinfo {author} {\bibfnamefont {X.-G.}\ \bibnamefont
  {{Wen}}},\ }\href@noop {} {\bibfield  {journal} {\bibinfo  {journal} {ArXiv
  e-prints}\ } (\bibinfo {year} {2012})},\ \Eprint
  {http://arxiv.org/abs/1212.4863} {arXiv:1212.4863 [cond-mat.str-el]}
  \BibitemShut {NoStop}%
\bibitem [{\citenamefont {Fidkowski}\ and\ \citenamefont
  {Kitaev}(2011)}]{Fidkowski2011}%
  \BibitemOpen
  \bibfield  {author} {\bibinfo {author} {\bibfnamefont {L.}~\bibnamefont
  {Fidkowski}}\ and\ \bibinfo {author} {\bibfnamefont {A.}~\bibnamefont
  {Kitaev}},\ }\href {\doibase 10.1103/PhysRevB.83.075103} {\bibfield
  {journal} {\bibinfo  {journal} {Phys. Rev. B}\ }\textbf {\bibinfo {volume}
  {83}},\ \bibinfo {pages} {075103} (\bibinfo {year} {2011})}\BibitemShut
  {NoStop}%
\bibitem [{\citenamefont {Qi}(2013)}]{Qi2013}%
  \BibitemOpen
  \bibfield  {author} {\bibinfo {author} {\bibfnamefont {X.-L.}\ \bibnamefont
  {Qi}},\ }\href {http://stacks.iop.org/1367-2630/15/i=6/a=065002} {\bibfield
  {journal} {\bibinfo  {journal} {New J. Phys.}\ }\textbf {\bibinfo {volume}
  {15}},\ \bibinfo {pages} {065002} (\bibinfo {year} {2013})}\BibitemShut
  {NoStop}%
\bibitem [{\citenamefont {Yao}\ and\ \citenamefont {Ryu}(2013)}]{Yao2013}%
  \BibitemOpen
  \bibfield  {author} {\bibinfo {author} {\bibfnamefont {H.}~\bibnamefont
  {Yao}}\ and\ \bibinfo {author} {\bibfnamefont {S.}~\bibnamefont {Ryu}},\
  }\href {\doibase 10.1103/PhysRevB.88.064507} {\bibfield  {journal} {\bibinfo
  {journal} {Phys. Rev. B}\ }\textbf {\bibinfo {volume} {88}},\ \bibinfo
  {pages} {064507} (\bibinfo {year} {2013})}\BibitemShut {NoStop}%
\bibitem [{\citenamefont {Tang}\ and\ \citenamefont {Wen}(2012)}]{Tang2012}%
  \BibitemOpen
  \bibfield  {author} {\bibinfo {author} {\bibfnamefont {E.}~\bibnamefont
  {Tang}}\ and\ \bibinfo {author} {\bibfnamefont {X.-G.}\ \bibnamefont {Wen}},\
  }\href {\doibase 10.1103/PhysRevLett.109.096403} {\bibfield  {journal}
  {\bibinfo  {journal} {Phys. Rev. Lett.}\ }\textbf {\bibinfo {volume} {109}},\
  \bibinfo {pages} {096403} (\bibinfo {year} {2012})}\BibitemShut {NoStop}%
\bibitem [{\citenamefont {Wu}\ \emph {et~al.}(2006)\citenamefont {Wu},
  \citenamefont {Bernevig},\ and\ \citenamefont {Zhang}}]{Wu2006}%
  \BibitemOpen
  \bibfield  {author} {\bibinfo {author} {\bibfnamefont {C.}~\bibnamefont
  {Wu}}, \bibinfo {author} {\bibfnamefont {B.~A.}\ \bibnamefont {Bernevig}}, \
  and\ \bibinfo {author} {\bibfnamefont {S.-C.}\ \bibnamefont {Zhang}},\ }\href
  {\doibase 10.1103/PhysRevLett.96.106401} {\bibfield  {journal} {\bibinfo
  {journal} {Phys. Rev. Lett.}\ }\textbf {\bibinfo {volume} {96}},\ \bibinfo
  {pages} {106401} (\bibinfo {year} {2006})}\BibitemShut {NoStop}%
\bibitem [{\citenamefont {Xu}\ and\ \citenamefont {Moore}(2006)}]{Xu2006}%
  \BibitemOpen
  \bibfield  {author} {\bibinfo {author} {\bibfnamefont {C.}~\bibnamefont
  {Xu}}\ and\ \bibinfo {author} {\bibfnamefont {J.~E.}\ \bibnamefont {Moore}},\
  }\href {\doibase 10.1103/PhysRevB.73.045322} {\bibfield  {journal} {\bibinfo
  {journal} {Phys. Rev. B}\ }\textbf {\bibinfo {volume} {73}},\ \bibinfo
  {pages} {045322} (\bibinfo {year} {2006})}\BibitemShut {NoStop}%
\bibitem [{\citenamefont {Zheng}\ \emph {et~al.}(2011)\citenamefont {Zheng},
  \citenamefont {Zhang},\ and\ \citenamefont {Wu}}]{Zheng2011}%
  \BibitemOpen
  \bibfield  {author} {\bibinfo {author} {\bibfnamefont {D.}~\bibnamefont
  {Zheng}}, \bibinfo {author} {\bibfnamefont {G.-M.}\ \bibnamefont {Zhang}}, \
  and\ \bibinfo {author} {\bibfnamefont {C.}~\bibnamefont {Wu}},\ }\href
  {\doibase 10.1103/PhysRevB.84.205121} {\bibfield  {journal} {\bibinfo
  {journal} {Phys. Rev. B}\ }\textbf {\bibinfo {volume} {84}},\ \bibinfo
  {pages} {205121} (\bibinfo {year} {2011})}\BibitemShut {NoStop}%
\bibitem [{\citenamefont {Hohenadler}\ \emph {et~al.}(2011)\citenamefont
  {Hohenadler}, \citenamefont {Lang},\ and\ \citenamefont
  {Assaad}}]{Hohenadler2011}%
  \BibitemOpen
  \bibfield  {author} {\bibinfo {author} {\bibfnamefont {M.}~\bibnamefont
  {Hohenadler}}, \bibinfo {author} {\bibfnamefont {T.~C.}\ \bibnamefont
  {Lang}}, \ and\ \bibinfo {author} {\bibfnamefont {F.~F.}\ \bibnamefont
  {Assaad}},\ }\href {\doibase 10.1103/PhysRevLett.106.100403} {\bibfield
  {journal} {\bibinfo  {journal} {Phys. Rev. Lett.}\ }\textbf {\bibinfo
  {volume} {106}},\ \bibinfo {pages} {100403} (\bibinfo {year}
  {2011})}\BibitemShut {NoStop}%
\bibitem [{\citenamefont {Yu}\ \emph {et~al.}(2011)\citenamefont {Yu},
  \citenamefont {Xie},\ and\ \citenamefont {Li}}]{Yu2011}%
  \BibitemOpen
  \bibfield  {author} {\bibinfo {author} {\bibfnamefont {S.-L.}\ \bibnamefont
  {Yu}}, \bibinfo {author} {\bibfnamefont {X.~C.}\ \bibnamefont {Xie}}, \ and\
  \bibinfo {author} {\bibfnamefont {J.-X.}\ \bibnamefont {Li}},\ }\href
  {\doibase 10.1103/PhysRevLett.107.010401} {\bibfield  {journal} {\bibinfo
  {journal} {Phys. Rev. Lett.}\ }\textbf {\bibinfo {volume} {107}},\ \bibinfo
  {pages} {010401} (\bibinfo {year} {2011})}\BibitemShut {NoStop}%
\bibitem [{\citenamefont {Raghu}\ \emph {et~al.}(2008)\citenamefont {Raghu},
  \citenamefont {Qi}, \citenamefont {Honerkamp},\ and\ \citenamefont
  {Zhang}}]{Raghu2008}%
  \BibitemOpen
  \bibfield  {author} {\bibinfo {author} {\bibfnamefont {S.}~\bibnamefont
  {Raghu}}, \bibinfo {author} {\bibfnamefont {X.-L.}\ \bibnamefont {Qi}},
  \bibinfo {author} {\bibfnamefont {C.}~\bibnamefont {Honerkamp}}, \ and\
  \bibinfo {author} {\bibfnamefont {S.-C.}\ \bibnamefont {Zhang}},\ }\href
  {\doibase 10.1103/PhysRevLett.100.156401} {\bibfield  {journal} {\bibinfo
  {journal} {Phys. Rev. Lett.}\ }\textbf {\bibinfo {volume} {100}},\ \bibinfo
  {pages} {156401} (\bibinfo {year} {2008})}\BibitemShut {NoStop}%
\bibitem [{\citenamefont {Dzero}\ \emph {et~al.}(2010)\citenamefont {Dzero},
  \citenamefont {Sun}, \citenamefont {Galitski},\ and\ \citenamefont
  {Coleman}}]{Dzero2010}%
  \BibitemOpen
  \bibfield  {author} {\bibinfo {author} {\bibfnamefont {M.}~\bibnamefont
  {Dzero}}, \bibinfo {author} {\bibfnamefont {K.}~\bibnamefont {Sun}}, \bibinfo
  {author} {\bibfnamefont {V.}~\bibnamefont {Galitski}}, \ and\ \bibinfo
  {author} {\bibfnamefont {P.}~\bibnamefont {Coleman}},\ }\href {\doibase
  10.1103/PhysRevLett.104.106408} {\bibfield  {journal} {\bibinfo  {journal}
  {Phys. Rev. Lett.}\ }\textbf {\bibinfo {volume} {104}},\ \bibinfo {pages}
  {106408} (\bibinfo {year} {2010})}\BibitemShut {NoStop}%
\bibitem [{\citenamefont {Shitade}\ \emph {et~al.}(2009)\citenamefont
  {Shitade}, \citenamefont {Katsura}, \citenamefont
  {Kune\ifmmode~\check{s}\else \v{s}\fi{}}, \citenamefont {Qi}, \citenamefont
  {Zhang},\ and\ \citenamefont {Nagaosa}}]{Shitade2009}%
  \BibitemOpen
  \bibfield  {author} {\bibinfo {author} {\bibfnamefont {A.}~\bibnamefont
  {Shitade}}, \bibinfo {author} {\bibfnamefont {H.}~\bibnamefont {Katsura}},
  \bibinfo {author} {\bibfnamefont {J.}~\bibnamefont
  {Kune\ifmmode~\check{s}\else \v{s}\fi{}}}, \bibinfo {author} {\bibfnamefont
  {X.-L.}\ \bibnamefont {Qi}}, \bibinfo {author} {\bibfnamefont {S.-C.}\
  \bibnamefont {Zhang}}, \ and\ \bibinfo {author} {\bibfnamefont
  {N.}~\bibnamefont {Nagaosa}},\ }\href {\doibase
  10.1103/PhysRevLett.102.256403} {\bibfield  {journal} {\bibinfo  {journal}
  {Phys. Rev. Lett.}\ }\textbf {\bibinfo {volume} {102}},\ \bibinfo {pages}
  {256403} (\bibinfo {year} {2009})}\BibitemShut {NoStop}%
\bibitem [{\citenamefont {Zhang}\ \emph
  {et~al.}(2012{\natexlab{a}})\citenamefont {Zhang}, \citenamefont {Zhang},
  \citenamefont {Wang}, \citenamefont {Felser},\ and\ \citenamefont
  {Zhang}}]{Zhang2012}%
  \BibitemOpen
  \bibfield  {author} {\bibinfo {author} {\bibfnamefont {X.}~\bibnamefont
  {Zhang}}, \bibinfo {author} {\bibfnamefont {H.}~\bibnamefont {Zhang}},
  \bibinfo {author} {\bibfnamefont {J.}~\bibnamefont {Wang}}, \bibinfo {author}
  {\bibfnamefont {C.}~\bibnamefont {Felser}}, \ and\ \bibinfo {author}
  {\bibfnamefont {S.-C.}\ \bibnamefont {Zhang}},\ }\href {\doibase
  10.1126/science.1216184} {\bibfield  {journal} {\bibinfo  {journal}
  {Science}\ }\textbf {\bibinfo {volume} {335}},\ \bibinfo {pages} {1464}
  (\bibinfo {year} {2012}{\natexlab{a}})}\BibitemShut {NoStop}%
\bibitem [{\citenamefont {Rachel}\ and\ \citenamefont
  {Le~Hur}(2010)}]{Rachel2010}%
  \BibitemOpen
  \bibfield  {author} {\bibinfo {author} {\bibfnamefont {S.}~\bibnamefont
  {Rachel}}\ and\ \bibinfo {author} {\bibfnamefont {K.}~\bibnamefont
  {Le~Hur}},\ }\href {\doibase 10.1103/PhysRevB.82.075106} {\bibfield
  {journal} {\bibinfo  {journal} {Phys. Rev. B}\ }\textbf {\bibinfo {volume}
  {82}},\ \bibinfo {pages} {075106} (\bibinfo {year} {2010})}\BibitemShut
  {NoStop}%
\bibitem [{\citenamefont {Varney}\ \emph {et~al.}(2010)\citenamefont {Varney},
  \citenamefont {Sun}, \citenamefont {Rigol},\ and\ \citenamefont
  {Galitski}}]{Varney2010}%
  \BibitemOpen
  \bibfield  {author} {\bibinfo {author} {\bibfnamefont {C.~N.}\ \bibnamefont
  {Varney}}, \bibinfo {author} {\bibfnamefont {K.}~\bibnamefont {Sun}},
  \bibinfo {author} {\bibfnamefont {M.}~\bibnamefont {Rigol}}, \ and\ \bibinfo
  {author} {\bibfnamefont {V.}~\bibnamefont {Galitski}},\ }\href {\doibase
  10.1103/PhysRevB.82.115125} {\bibfield  {journal} {\bibinfo  {journal} {Phys.
  Rev. B}\ }\textbf {\bibinfo {volume} {82}},\ \bibinfo {pages} {115125}
  (\bibinfo {year} {2010})}\BibitemShut {NoStop}%
\bibitem [{\citenamefont {Yuan}\ \emph {et~al.}(2012)\citenamefont {Yuan},
  \citenamefont {Gao}, \citenamefont {Chen}, \citenamefont {Ye}, \citenamefont
  {Zhou},\ and\ \citenamefont {Zhang}}]{Yuan2012}%
  \BibitemOpen
  \bibfield  {author} {\bibinfo {author} {\bibfnamefont {J.}~\bibnamefont
  {Yuan}}, \bibinfo {author} {\bibfnamefont {J.-H.}\ \bibnamefont {Gao}},
  \bibinfo {author} {\bibfnamefont {W.-Q.}\ \bibnamefont {Chen}}, \bibinfo
  {author} {\bibfnamefont {F.}~\bibnamefont {Ye}}, \bibinfo {author}
  {\bibfnamefont {Y.}~\bibnamefont {Zhou}}, \ and\ \bibinfo {author}
  {\bibfnamefont {F.-C.}\ \bibnamefont {Zhang}},\ }\href {\doibase
  10.1103/PhysRevB.86.104505} {\bibfield  {journal} {\bibinfo  {journal} {Phys.
  Rev. B}\ }\textbf {\bibinfo {volume} {86}},\ \bibinfo {pages} {104505}
  (\bibinfo {year} {2012})}\BibitemShut {NoStop}%
\bibitem [{\citenamefont {Chen}\ \emph {et~al.}(2011)\citenamefont {Chen},
  \citenamefont {Liu},\ and\ \citenamefont {Wen}}]{Chen2011}%
  \BibitemOpen
  \bibfield  {author} {\bibinfo {author} {\bibfnamefont {X.}~\bibnamefont
  {Chen}}, \bibinfo {author} {\bibfnamefont {Z.-X.}\ \bibnamefont {Liu}}, \
  and\ \bibinfo {author} {\bibfnamefont {X.-G.}\ \bibnamefont {Wen}},\ }\href
  {\doibase 10.1103/PhysRevB.84.235141} {\bibfield  {journal} {\bibinfo
  {journal} {Phys. Rev. B}\ }\textbf {\bibinfo {volume} {84}},\ \bibinfo
  {pages} {235141} (\bibinfo {year} {2011})}\BibitemShut {NoStop}%
\bibitem [{\citenamefont {Turner}\ \emph {et~al.}(2011)\citenamefont {Turner},
  \citenamefont {Pollmann},\ and\ \citenamefont {Berg}}]{Turner2011}%
  \BibitemOpen
  \bibfield  {author} {\bibinfo {author} {\bibfnamefont {A.~M.}\ \bibnamefont
  {Turner}}, \bibinfo {author} {\bibfnamefont {F.}~\bibnamefont {Pollmann}}, \
  and\ \bibinfo {author} {\bibfnamefont {E.}~\bibnamefont {Berg}},\ }\href
  {\doibase 10.1103/PhysRevB.83.075102} {\bibfield  {journal} {\bibinfo
  {journal} {Phys. Rev. B}\ }\textbf {\bibinfo {volume} {83}},\ \bibinfo
  {pages} {075102} (\bibinfo {year} {2011})}\BibitemShut {NoStop}%
\bibitem [{\citenamefont {Lu}\ and\ \citenamefont {Vishwanath}(2012)}]{Lu2012}%
  \BibitemOpen
  \bibfield  {author} {\bibinfo {author} {\bibfnamefont {Y.-M.}\ \bibnamefont
  {Lu}}\ and\ \bibinfo {author} {\bibfnamefont {A.}~\bibnamefont
  {Vishwanath}},\ }\href {\doibase 10.1103/PhysRevB.86.125119} {\bibfield
  {journal} {\bibinfo  {journal} {Phys. Rev. B}\ }\textbf {\bibinfo {volume}
  {86}},\ \bibinfo {pages} {125119} (\bibinfo {year} {2012})}\BibitemShut
  {NoStop}%
\bibitem [{\citenamefont {{Gu}}\ and\ \citenamefont {{Wen}}(2012)}]{Gu2012}%
  \BibitemOpen
  \bibfield  {author} {\bibinfo {author} {\bibfnamefont {Z.-C.}\ \bibnamefont
  {{Gu}}}\ and\ \bibinfo {author} {\bibfnamefont {X.-G.}\ \bibnamefont
  {{Wen}}},\ }\href@noop {} {\  (\bibinfo {year} {2012})},\ \Eprint
  {http://arxiv.org/abs/1201.2648} {arXiv:1201.2648} \BibitemShut {NoStop}%
\bibitem [{\citenamefont {Wang}\ \emph {et~al.}(2014)\citenamefont {Wang},
  \citenamefont {Potter},\ and\ \citenamefont {Senthil}}]{Wang2013}%
  \BibitemOpen
  \bibfield  {author} {\bibinfo {author} {\bibfnamefont {C.}~\bibnamefont
  {Wang}}, \bibinfo {author} {\bibfnamefont {A.~C.}\ \bibnamefont {Potter}}, \
  and\ \bibinfo {author} {\bibfnamefont {T.}~\bibnamefont {Senthil}},\ }\href
  {\doibase 10.1126/science.1243326} {\bibfield  {journal} {\bibinfo  {journal}
  {Science}\ }\textbf {\bibinfo {volume} {343}},\ \bibinfo {pages} {629}
  (\bibinfo {year} {2014})}\BibitemShut {NoStop}%
\bibitem [{\citenamefont {Wang}\ \emph {et~al.}(2010)\citenamefont {Wang},
  \citenamefont {Qi},\ and\ \citenamefont {Zhang}}]{Wang2010}%
  \BibitemOpen
  \bibfield  {author} {\bibinfo {author} {\bibfnamefont {Z.}~\bibnamefont
  {Wang}}, \bibinfo {author} {\bibfnamefont {X.-L.}\ \bibnamefont {Qi}}, \ and\
  \bibinfo {author} {\bibfnamefont {S.-C.}\ \bibnamefont {Zhang}},\ }\href
  {\doibase 10.1103/PhysRevLett.105.256803} {\bibfield  {journal} {\bibinfo
  {journal} {Phys. Rev. Lett.}\ }\textbf {\bibinfo {volume} {105}},\ \bibinfo
  {pages} {256803} (\bibinfo {year} {2010})}\BibitemShut {NoStop}%
\bibitem [{\citenamefont {Wang}\ \emph {et~al.}(2011)\citenamefont {Wang},
  \citenamefont {Dai},\ and\ \citenamefont {Xie}}]{Wang2011}%
  \BibitemOpen
  \bibfield  {author} {\bibinfo {author} {\bibfnamefont {L.}~\bibnamefont
  {Wang}}, \bibinfo {author} {\bibfnamefont {X.}~\bibnamefont {Dai}}, \ and\
  \bibinfo {author} {\bibfnamefont {X.~C.}\ \bibnamefont {Xie}},\ }\href
  {\doibase 10.1103/PhysRevB.84.205116} {\bibfield  {journal} {\bibinfo
  {journal} {Phys. Rev. B}\ }\textbf {\bibinfo {volume} {84}},\ \bibinfo
  {pages} {205116} (\bibinfo {year} {2011})}\BibitemShut {NoStop}%
\bibitem [{\citenamefont {Wang}\ and\ \citenamefont {Zhang}(2012)}]{Wang2012}%
  \BibitemOpen
  \bibfield  {author} {\bibinfo {author} {\bibfnamefont {Z.}~\bibnamefont
  {Wang}}\ and\ \bibinfo {author} {\bibfnamefont {S.-C.}\ \bibnamefont
  {Zhang}},\ }\href {\doibase 10.1103/PhysRevX.2.031008} {\bibfield  {journal}
  {\bibinfo  {journal} {Phys. Rev. X}\ }\textbf {\bibinfo {volume} {2}},\
  \bibinfo {pages} {031008} (\bibinfo {year} {2012})}\BibitemShut {NoStop}%
\bibitem [{\citenamefont {Manmana}\ \emph {et~al.}(2012)\citenamefont
  {Manmana}, \citenamefont {Essin}, \citenamefont {Noack},\ and\ \citenamefont
  {Gurarie}}]{Manmana2012}%
  \BibitemOpen
  \bibfield  {author} {\bibinfo {author} {\bibfnamefont {S.~R.}\ \bibnamefont
  {Manmana}}, \bibinfo {author} {\bibfnamefont {A.~M.}\ \bibnamefont {Essin}},
  \bibinfo {author} {\bibfnamefont {R.~M.}\ \bibnamefont {Noack}}, \ and\
  \bibinfo {author} {\bibfnamefont {V.}~\bibnamefont {Gurarie}},\ }\href
  {\doibase 10.1103/PhysRevB.86.205119} {\bibfield  {journal} {\bibinfo
  {journal} {Phys. Rev. B}\ }\textbf {\bibinfo {volume} {86}},\ \bibinfo
  {pages} {205119} (\bibinfo {year} {2012})}\BibitemShut {NoStop}%
\bibitem [{\citenamefont {Wang}\ \emph {et~al.}(2012)\citenamefont {Wang},
  \citenamefont {Jiang}, \citenamefont {Dai},\ and\ \citenamefont
  {Xie}}]{Wang2012a}%
  \BibitemOpen
  \bibfield  {author} {\bibinfo {author} {\bibfnamefont {L.}~\bibnamefont
  {Wang}}, \bibinfo {author} {\bibfnamefont {H.}~\bibnamefont {Jiang}},
  \bibinfo {author} {\bibfnamefont {X.}~\bibnamefont {Dai}}, \ and\ \bibinfo
  {author} {\bibfnamefont {X.~C.}\ \bibnamefont {Xie}},\ }\href {\doibase
  10.1103/PhysRevB.85.235135} {\bibfield  {journal} {\bibinfo  {journal} {Phys.
  Rev. B}\ }\textbf {\bibinfo {volume} {85}},\ \bibinfo {pages} {235135}
  (\bibinfo {year} {2012})}\BibitemShut {NoStop}%
\bibitem [{\citenamefont {Ara\'ujo}\ \emph {et~al.}(2013)\citenamefont
  {Ara\'ujo}, \citenamefont {Castro},\ and\ \citenamefont
  {Sacramento}}]{Araujo2013}%
  \BibitemOpen
  \bibfield  {author} {\bibinfo {author} {\bibfnamefont {M.~A.~N.}\
  \bibnamefont {Ara\'ujo}}, \bibinfo {author} {\bibfnamefont {E.~V.}\
  \bibnamefont {Castro}}, \ and\ \bibinfo {author} {\bibfnamefont {P.~D.}\
  \bibnamefont {Sacramento}},\ }\href {\doibase 10.1103/PhysRevB.87.085109}
  {\bibfield  {journal} {\bibinfo  {journal} {Phys. Rev. B}\ }\textbf {\bibinfo
  {volume} {87}},\ \bibinfo {pages} {085109} (\bibinfo {year}
  {2013})}\BibitemShut {NoStop}%
\bibitem [{\citenamefont {Hung}\ \emph {et~al.}(2013)\citenamefont {Hung},
  \citenamefont {Wang}, \citenamefont {Gu},\ and\ \citenamefont
  {Fiete}}]{Hung2013}%
  \BibitemOpen
  \bibfield  {author} {\bibinfo {author} {\bibfnamefont {H.-H.}\ \bibnamefont
  {Hung}}, \bibinfo {author} {\bibfnamefont {L.}~\bibnamefont {Wang}}, \bibinfo
  {author} {\bibfnamefont {Z.-C.}\ \bibnamefont {Gu}}, \ and\ \bibinfo {author}
  {\bibfnamefont {G.~A.}\ \bibnamefont {Fiete}},\ }\href {\doibase
  10.1103/PhysRevB.87.121113} {\bibfield  {journal} {\bibinfo  {journal} {Phys.
  Rev. B}\ }\textbf {\bibinfo {volume} {87}},\ \bibinfo {pages} {121113}
  (\bibinfo {year} {2013})}\BibitemShut {NoStop}%
\bibitem [{\citenamefont {Lang}\ \emph {et~al.}(2013)\citenamefont {Lang},
  \citenamefont {Essin}, \citenamefont {Gurarie},\ and\ \citenamefont
  {Wessel}}]{Lang2013}%
  \BibitemOpen
  \bibfield  {author} {\bibinfo {author} {\bibfnamefont {T.~C.}\ \bibnamefont
  {Lang}}, \bibinfo {author} {\bibfnamefont {A.~M.}\ \bibnamefont {Essin}},
  \bibinfo {author} {\bibfnamefont {V.}~\bibnamefont {Gurarie}}, \ and\
  \bibinfo {author} {\bibfnamefont {S.}~\bibnamefont {Wessel}},\ }\href
  {\doibase 10.1103/PhysRevB.87.205101} {\bibfield  {journal} {\bibinfo
  {journal} {Phys. Rev. B}\ }\textbf {\bibinfo {volume} {87}},\ \bibinfo
  {pages} {205101} (\bibinfo {year} {2013})}\BibitemShut {NoStop}%
\bibitem [{\citenamefont {Amico}\ \emph {et~al.}(2008)\citenamefont {Amico},
  \citenamefont {Fazio}, \citenamefont {Osterloh},\ and\ \citenamefont
  {Vedral}}]{Amico2008}%
  \BibitemOpen
  \bibfield  {author} {\bibinfo {author} {\bibfnamefont {L.}~\bibnamefont
  {Amico}}, \bibinfo {author} {\bibfnamefont {R.}~\bibnamefont {Fazio}},
  \bibinfo {author} {\bibfnamefont {A.}~\bibnamefont {Osterloh}}, \ and\
  \bibinfo {author} {\bibfnamefont {V.}~\bibnamefont {Vedral}},\ }\href
  {\doibase 10.1103/RevModPhys.80.517} {\bibfield  {journal} {\bibinfo
  {journal} {Rev. Mod. Phys.}\ }\textbf {\bibinfo {volume} {80}},\ \bibinfo
  {pages} {517} (\bibinfo {year} {2008})}\BibitemShut {NoStop}%
\bibitem [{\citenamefont {Eisert}\ \emph {et~al.}(2010)\citenamefont {Eisert},
  \citenamefont {Cramer},\ and\ \citenamefont {Plenio}}]{Eisert2010}%
  \BibitemOpen
  \bibfield  {author} {\bibinfo {author} {\bibfnamefont {J.}~\bibnamefont
  {Eisert}}, \bibinfo {author} {\bibfnamefont {M.}~\bibnamefont {Cramer}}, \
  and\ \bibinfo {author} {\bibfnamefont {M.~B.}\ \bibnamefont {Plenio}},\
  }\href {\doibase 10.1103/RevModPhys.82.277} {\bibfield  {journal} {\bibinfo
  {journal} {Rev. Mod. Phys.}\ }\textbf {\bibinfo {volume} {82}},\ \bibinfo
  {pages} {277} (\bibinfo {year} {2010})}\BibitemShut {NoStop}%
\bibitem [{\citenamefont {Vidal}\ \emph {et~al.}(2003)\citenamefont {Vidal},
  \citenamefont {Latorre}, \citenamefont {Rico},\ and\ \citenamefont
  {Kitaev}}]{Vidal2003}%
  \BibitemOpen
  \bibfield  {author} {\bibinfo {author} {\bibfnamefont {G.}~\bibnamefont
  {Vidal}}, \bibinfo {author} {\bibfnamefont {J.~I.}\ \bibnamefont {Latorre}},
  \bibinfo {author} {\bibfnamefont {E.}~\bibnamefont {Rico}}, \ and\ \bibinfo
  {author} {\bibfnamefont {A.}~\bibnamefont {Kitaev}},\ }\href {\doibase
  10.1103/PhysRevLett.90.227902} {\bibfield  {journal} {\bibinfo  {journal}
  {Phys. Rev. Lett.}\ }\textbf {\bibinfo {volume} {90}},\ \bibinfo {pages}
  {227902} (\bibinfo {year} {2003})}\BibitemShut {NoStop}%
\bibitem [{\citenamefont {Calabrese}\ and\ \citenamefont
  {Cardy}(2004)}]{Calabrese2004}%
  \BibitemOpen
  \bibfield  {author} {\bibinfo {author} {\bibfnamefont {P.}~\bibnamefont
  {Calabrese}}\ and\ \bibinfo {author} {\bibfnamefont {J.}~\bibnamefont
  {Cardy}},\ }\href {http://stacks.iop.org/1742-5468/2004/i=06/a=P06002}
  {\bibfield  {journal} {\bibinfo  {journal} {J. Stat. Mech: Theory Exp.}\
  }\textbf {\bibinfo {volume} {2004}},\ \bibinfo {pages} {P06002} (\bibinfo
  {year} {2004})}\BibitemShut {NoStop}%
\bibitem [{\citenamefont {Kitaev}\ and\ \citenamefont
  {Preskill}(2006)}]{Kitaev2006}%
  \BibitemOpen
  \bibfield  {author} {\bibinfo {author} {\bibfnamefont {A.}~\bibnamefont
  {Kitaev}}\ and\ \bibinfo {author} {\bibfnamefont {J.}~\bibnamefont
  {Preskill}},\ }\href {\doibase 10.1103/PhysRevLett.96.110404} {\bibfield
  {journal} {\bibinfo  {journal} {Phys. Rev. Lett.}\ }\textbf {\bibinfo
  {volume} {96}},\ \bibinfo {pages} {110404} (\bibinfo {year}
  {2006})}\BibitemShut {NoStop}%
\bibitem [{\citenamefont {Levin}\ and\ \citenamefont {Wen}(2006)}]{Levin2006}%
  \BibitemOpen
  \bibfield  {author} {\bibinfo {author} {\bibfnamefont {M.}~\bibnamefont
  {Levin}}\ and\ \bibinfo {author} {\bibfnamefont {X.-G.}\ \bibnamefont
  {Wen}},\ }\href {\doibase 10.1103/PhysRevLett.96.110405} {\bibfield
  {journal} {\bibinfo  {journal} {Phys. Rev. Lett.}\ }\textbf {\bibinfo
  {volume} {96}},\ \bibinfo {pages} {110405} (\bibinfo {year}
  {2006})}\BibitemShut {NoStop}%
\bibitem [{\citenamefont {Zhang}\ \emph {et~al.}(2011)\citenamefont {Zhang},
  \citenamefont {Grover},\ and\ \citenamefont {Vishwanath}}]{Zhang2011}%
  \BibitemOpen
  \bibfield  {author} {\bibinfo {author} {\bibfnamefont {Y.}~\bibnamefont
  {Zhang}}, \bibinfo {author} {\bibfnamefont {T.}~\bibnamefont {Grover}}, \
  and\ \bibinfo {author} {\bibfnamefont {A.}~\bibnamefont {Vishwanath}},\
  }\href {\doibase 10.1103/PhysRevLett.107.067202} {\bibfield  {journal}
  {\bibinfo  {journal} {Phys. Rev. Lett.}\ }\textbf {\bibinfo {volume} {107}},\
  \bibinfo {pages} {067202} (\bibinfo {year} {2011})}\BibitemShut {NoStop}%
\bibitem [{\citenamefont {Zhang}\ \emph
  {et~al.}(2012{\natexlab{b}})\citenamefont {Zhang}, \citenamefont {Grover},
  \citenamefont {Turner}, \citenamefont {Oshikawa},\ and\ \citenamefont
  {Vishwanath}}]{Zhang2012a}%
  \BibitemOpen
  \bibfield  {author} {\bibinfo {author} {\bibfnamefont {Y.}~\bibnamefont
  {Zhang}}, \bibinfo {author} {\bibfnamefont {T.}~\bibnamefont {Grover}},
  \bibinfo {author} {\bibfnamefont {A.}~\bibnamefont {Turner}}, \bibinfo
  {author} {\bibfnamefont {M.}~\bibnamefont {Oshikawa}}, \ and\ \bibinfo
  {author} {\bibfnamefont {A.}~\bibnamefont {Vishwanath}},\ }\href {\doibase
  10.1103/PhysRevB.85.235151} {\bibfield  {journal} {\bibinfo  {journal} {Phys.
  Rev. B}\ }\textbf {\bibinfo {volume} {85}},\ \bibinfo {pages} {235151}
  (\bibinfo {year} {2012}{\natexlab{b}})}\BibitemShut {NoStop}%
\bibitem [{\citenamefont {Yan}\ \emph {et~al.}(2011)\citenamefont {Yan},
  \citenamefont {Huse},\ and\ \citenamefont {White}}]{Yan2011}%
  \BibitemOpen
  \bibfield  {author} {\bibinfo {author} {\bibfnamefont {S.}~\bibnamefont
  {Yan}}, \bibinfo {author} {\bibfnamefont {D.~A.}\ \bibnamefont {Huse}}, \
  and\ \bibinfo {author} {\bibfnamefont {S.~R.}\ \bibnamefont {White}},\ }\href
  {\doibase 10.1126/science.1201080} {\bibfield  {journal} {\bibinfo  {journal}
  {Science}\ }\textbf {\bibinfo {volume} {332}},\ \bibinfo {pages} {1173}
  (\bibinfo {year} {2011})}\BibitemShut {NoStop}%
\bibitem [{\citenamefont {Jiang}\ \emph
  {et~al.}(2012{\natexlab{a}})\citenamefont {Jiang}, \citenamefont {Wang},\
  and\ \citenamefont {Balents}}]{Jiang2012}%
  \BibitemOpen
  \bibfield  {author} {\bibinfo {author} {\bibfnamefont {H.-C.}\ \bibnamefont
  {Jiang}}, \bibinfo {author} {\bibfnamefont {Z.}~\bibnamefont {Wang}}, \ and\
  \bibinfo {author} {\bibfnamefont {L.}~\bibnamefont {Balents}},\ }\href
  {http://dx.doi.org/10.1038/nphys2465} {\bibfield  {journal} {\bibinfo
  {journal} {Nat. Phys.}\ }\textbf {\bibinfo {volume} {8}},\ \bibinfo {pages}
  {902} (\bibinfo {year} {2012}{\natexlab{a}})}\BibitemShut {NoStop}%
\bibitem [{\citenamefont {Jiang}\ \emph
  {et~al.}(2012{\natexlab{b}})\citenamefont {Jiang}, \citenamefont {Yao},\ and\
  \citenamefont {Balents}}]{Jiang2012a}%
  \BibitemOpen
  \bibfield  {author} {\bibinfo {author} {\bibfnamefont {H.-C.}\ \bibnamefont
  {Jiang}}, \bibinfo {author} {\bibfnamefont {H.}~\bibnamefont {Yao}}, \ and\
  \bibinfo {author} {\bibfnamefont {L.}~\bibnamefont {Balents}},\ }\href
  {\doibase 10.1103/PhysRevB.86.024424} {\bibfield  {journal} {\bibinfo
  {journal} {Phys. Rev. B}\ }\textbf {\bibinfo {volume} {86}},\ \bibinfo
  {pages} {024424} (\bibinfo {year} {2012}{\natexlab{b}})}\BibitemShut
  {NoStop}%
\bibitem [{\citenamefont {Li}\ and\ \citenamefont {Haldane}(2008)}]{Li2008}%
  \BibitemOpen
  \bibfield  {author} {\bibinfo {author} {\bibfnamefont {H.}~\bibnamefont
  {Li}}\ and\ \bibinfo {author} {\bibfnamefont {F.~D.~M.}\ \bibnamefont
  {Haldane}},\ }\href {\doibase 10.1103/PhysRevLett.101.010504} {\bibfield
  {journal} {\bibinfo  {journal} {Phys. Rev. Lett.}\ }\textbf {\bibinfo
  {volume} {101}},\ \bibinfo {pages} {010504} (\bibinfo {year}
  {2008})}\BibitemShut {NoStop}%
\bibitem [{\citenamefont {Ryu}\ and\ \citenamefont {Hatsugai}(2006)}]{Ryu2006}%
  \BibitemOpen
  \bibfield  {author} {\bibinfo {author} {\bibfnamefont {S.}~\bibnamefont
  {Ryu}}\ and\ \bibinfo {author} {\bibfnamefont {Y.}~\bibnamefont {Hatsugai}},\
  }\href {\doibase 10.1103/PhysRevB.73.245115} {\bibfield  {journal} {\bibinfo
  {journal} {Phys. Rev. B}\ }\textbf {\bibinfo {volume} {73}},\ \bibinfo
  {pages} {245115} (\bibinfo {year} {2006})}\BibitemShut {NoStop}%
\bibitem [{\citenamefont {Fidkowski}(2010)}]{Fidkowski2010}%
  \BibitemOpen
  \bibfield  {author} {\bibinfo {author} {\bibfnamefont {L.}~\bibnamefont
  {Fidkowski}},\ }\href {\doibase 10.1103/PhysRevLett.104.130502} {\bibfield
  {journal} {\bibinfo  {journal} {Phys. Rev. Lett.}\ }\textbf {\bibinfo
  {volume} {104}},\ \bibinfo {pages} {130502} (\bibinfo {year}
  {2010})}\BibitemShut {NoStop}%
\bibitem [{\citenamefont {Grover}(2013)}]{Grover2013}%
  \BibitemOpen
  \bibfield  {author} {\bibinfo {author} {\bibfnamefont {T.}~\bibnamefont
  {Grover}},\ }\href {\doibase 10.1103/PhysRevLett.111.130402} {\bibfield
  {journal} {\bibinfo  {journal} {Phys. Rev. Lett.}\ }\textbf {\bibinfo
  {volume} {111}},\ \bibinfo {pages} {130402} (\bibinfo {year}
  {2013})}\BibitemShut {NoStop}%
\bibitem [{\citenamefont {Assaad}\ \emph {et~al.}(2014)\citenamefont {Assaad},
  \citenamefont {Lang},\ and\ \citenamefont {Parisen~Toldin}}]{Assaad2013}%
  \BibitemOpen
  \bibfield  {author} {\bibinfo {author} {\bibfnamefont {F.~F.}\ \bibnamefont
  {Assaad}}, \bibinfo {author} {\bibfnamefont {T.~C.}\ \bibnamefont {Lang}}, \
  and\ \bibinfo {author} {\bibfnamefont {F.}~\bibnamefont {Parisen~Toldin}},\
  }\href {\doibase 10.1103/PhysRevB.89.125121} {\bibfield  {journal} {\bibinfo
  {journal} {Phys. Rev. B}\ }\textbf {\bibinfo {volume} {89}},\ \bibinfo
  {pages} {125121} (\bibinfo {year} {2014})}\BibitemShut {NoStop}%
\bibitem [{\citenamefont {Su}\ \emph {et~al.}(1979)\citenamefont {Su},
  \citenamefont {Schrieffer},\ and\ \citenamefont {Heeger}}]{Su1979}%
  \BibitemOpen
  \bibfield  {author} {\bibinfo {author} {\bibfnamefont {W.~P.}\ \bibnamefont
  {Su}}, \bibinfo {author} {\bibfnamefont {J.~R.}\ \bibnamefont {Schrieffer}},
  \ and\ \bibinfo {author} {\bibfnamefont {A.~J.}\ \bibnamefont {Heeger}},\
  }\href {\doibase 10.1103/PhysRevLett.42.1698} {\bibfield  {journal} {\bibinfo
   {journal} {Phys. Rev. Lett.}\ }\textbf {\bibinfo {volume} {42}},\ \bibinfo
  {pages} {1698} (\bibinfo {year} {1979})}\BibitemShut {NoStop}%
\bibitem [{\citenamefont {Kane}\ and\ \citenamefont {Mele}(2005)}]{Kane2005}%
  \BibitemOpen
  \bibfield  {author} {\bibinfo {author} {\bibfnamefont {C.~L.}\ \bibnamefont
  {Kane}}\ and\ \bibinfo {author} {\bibfnamefont {E.~J.}\ \bibnamefont
  {Mele}},\ }\href {\doibase 10.1103/PhysRevLett.95.146802} {\bibfield
  {journal} {\bibinfo  {journal} {Phys. Rev. Lett.}\ }\textbf {\bibinfo
  {volume} {95}},\ \bibinfo {pages} {146802} (\bibinfo {year}
  {2005})}\BibitemShut {NoStop}%
\bibitem [{\citenamefont {Peschel}(2003)}]{Peschel2003}%
  \BibitemOpen
  \bibfield  {author} {\bibinfo {author} {\bibfnamefont {I.}~\bibnamefont
  {Peschel}},\ }\href {http://stacks.iop.org/0305-4470/36/i=14/a=101}
  {\bibfield  {journal} {\bibinfo  {journal} {J. Phys. A: Math. Gen.}\ }\textbf
  {\bibinfo {volume} {36}},\ \bibinfo {pages} {L205} (\bibinfo {year}
  {2003})}\BibitemShut {NoStop}%
\bibitem [{\citenamefont {Cheong}\ and\ \citenamefont
  {Henley}(2004)}]{Cheong2004}%
  \BibitemOpen
  \bibfield  {author} {\bibinfo {author} {\bibfnamefont {S.-A.}\ \bibnamefont
  {Cheong}}\ and\ \bibinfo {author} {\bibfnamefont {C.~L.}\ \bibnamefont
  {Henley}},\ }\href {\doibase 10.1103/PhysRevB.69.075111} {\bibfield
  {journal} {\bibinfo  {journal} {Phys. Rev. B}\ }\textbf {\bibinfo {volume}
  {69}},\ \bibinfo {pages} {075111} (\bibinfo {year} {2004})}\BibitemShut
  {NoStop}%
\bibitem [{\citenamefont {{Kim}}(2013)}]{Kim2013a}%
  \BibitemOpen
  \bibfield  {author} {\bibinfo {author} {\bibfnamefont {I.~H.}\ \bibnamefont
  {{Kim}}},\ }\href@noop {} {\bibfield  {journal} {\bibinfo  {journal} {ArXiv
  e-prints}\ } (\bibinfo {year} {2013})},\ \Eprint
  {http://arxiv.org/abs/1306.4771} {arXiv:1306.4771} \BibitemShut {NoStop}%
\bibitem [{\citenamefont {Assaad}\ and\ \citenamefont
  {Evertz}(2008)}]{Assaad2008}%
  \BibitemOpen
  \bibfield  {author} {\bibinfo {author} {\bibfnamefont {F.~F.}\ \bibnamefont
  {Assaad}}\ and\ \bibinfo {author} {\bibfnamefont {H.~G.}\ \bibnamefont
  {Evertz}},\ }\href@noop {} {\emph {\bibinfo {title} {Computational
  Many-Particle Physics}}}\ (\bibinfo  {publisher} {Macmillan Publishers
  Limited. All rights reserved},\ \bibinfo {year} {2008})\ pp.\ \bibinfo
  {pages} {277--356}\BibitemShut {NoStop}%
\bibitem [{\citenamefont {Wang}\ \emph {et~al.}(2013)\citenamefont {Wang},
  \citenamefont {Li},\ and\ \citenamefont {Cho}}]{Wang2013b}%
  \BibitemOpen
  \bibfield  {author} {\bibinfo {author} {\bibfnamefont {H.~T.}\ \bibnamefont
  {Wang}}, \bibinfo {author} {\bibfnamefont {B.}~\bibnamefont {Li}}, \ and\
  \bibinfo {author} {\bibfnamefont {S.~Y.}\ \bibnamefont {Cho}},\ }\href
  {\doibase 10.1103/PhysRevB.87.054402} {\bibfield  {journal} {\bibinfo
  {journal} {Phys. Rev. B}\ }\textbf {\bibinfo {volume} {87}},\ \bibinfo
  {pages} {054402} (\bibinfo {year} {2013})}\BibitemShut {NoStop}%
\bibitem [{\citenamefont {des Cloizeaux}\ and\ \citenamefont
  {Pearson}(1962)}]{Cloizeaux1962}%
  \BibitemOpen
  \bibfield  {author} {\bibinfo {author} {\bibfnamefont {J.}~\bibnamefont {des
  Cloizeaux}}\ and\ \bibinfo {author} {\bibfnamefont {J.~J.}\ \bibnamefont
  {Pearson}},\ }\href {\doibase 10.1103/PhysRev.128.2131} {\bibfield  {journal}
  {\bibinfo  {journal} {Phys. Rev.}\ }\textbf {\bibinfo {volume} {128}},\
  \bibinfo {pages} {2131} (\bibinfo {year} {1962})}\BibitemShut {NoStop}%
\bibitem [{\citenamefont {Haldane}(1983)}]{Haldane1983}%
  \BibitemOpen
  \bibfield  {author} {\bibinfo {author} {\bibfnamefont {F.~D.~M.}\
  \bibnamefont {Haldane}},\ }\href {\doibase 10.1103/PhysRevLett.50.1153}
  {\bibfield  {journal} {\bibinfo  {journal} {Phys. Rev. Lett.}\ }\textbf
  {\bibinfo {volume} {50}},\ \bibinfo {pages} {1153} (\bibinfo {year}
  {1983})}\BibitemShut {NoStop}%
\bibitem [{\citenamefont {Yao}\ and\ \citenamefont {Qi}(2010)}]{Yao2010}%
  \BibitemOpen
  \bibfield  {author} {\bibinfo {author} {\bibfnamefont {H.}~\bibnamefont
  {Yao}}\ and\ \bibinfo {author} {\bibfnamefont {X.-L.}\ \bibnamefont {Qi}},\
  }\href {\doibase 10.1103/PhysRevLett.105.080501} {\bibfield  {journal}
  {\bibinfo  {journal} {Phys. Rev. Lett.}\ }\textbf {\bibinfo {volume} {105}},\
  \bibinfo {pages} {080501} (\bibinfo {year} {2010})}\BibitemShut {NoStop}%
\bibitem [{\citenamefont {{Oliveira}}\ \emph {et~al.}(2013)\citenamefont
  {{Oliveira}}, \citenamefont {{Ribeiro}},\ and\ \citenamefont
  {{Sacramento}}}]{Oliveira2013}%
  \BibitemOpen
  \bibfield  {author} {\bibinfo {author} {\bibfnamefont {T.~P.}\ \bibnamefont
  {{Oliveira}}}, \bibinfo {author} {\bibfnamefont {P.}~\bibnamefont
  {{Ribeiro}}}, \ and\ \bibinfo {author} {\bibfnamefont {P.~D.}\ \bibnamefont
  {{Sacramento}}},\ }\href@noop {} {\bibfield  {journal} {\bibinfo  {journal}
  {ArXiv e-prints}\ } (\bibinfo {year} {2013})},\ \Eprint
  {http://arxiv.org/abs/1312.7782} {arXiv:1312.7782} \BibitemShut {NoStop}%
\bibitem [{\citenamefont {{Broecker}}\ and\ \citenamefont
  {{Trebst}}(2014)}]{Broecker2014}%
  \BibitemOpen
  \bibfield  {author} {\bibinfo {author} {\bibfnamefont {P.}~\bibnamefont
  {{Broecker}}}\ and\ \bibinfo {author} {\bibfnamefont {S.}~\bibnamefont
  {{Trebst}}},\ }\href@noop {} {\bibfield  {journal} {\bibinfo  {journal}
  {ArXiv e-prints}\ } (\bibinfo {year} {2014})},\ \Eprint
  {http://arxiv.org/abs/1404.3027} {arXiv:1404.3027} \BibitemShut {NoStop}%
\bibitem [{\citenamefont {Sfiligoi}\ \emph {et~al.}(2009)\citenamefont
  {Sfiligoi}, \citenamefont {Bradley}, \citenamefont {Holzman}, \citenamefont
  {Mhashilkar}, \citenamefont {Padhi},\ and\ \citenamefont
  {Wurthwein}}]{Sfiligoi2009}%
  \BibitemOpen
  \bibfield  {author} {\bibinfo {author} {\bibfnamefont {I.}~\bibnamefont
  {Sfiligoi}}, \bibinfo {author} {\bibfnamefont {D.}~\bibnamefont {Bradley}},
  \bibinfo {author} {\bibfnamefont {B.}~\bibnamefont {Holzman}}, \bibinfo
  {author} {\bibfnamefont {P.}~\bibnamefont {Mhashilkar}}, \bibinfo {author}
  {\bibfnamefont {S.}~\bibnamefont {Padhi}}, \ and\ \bibinfo {author}
  {\bibfnamefont {F.}~\bibnamefont {Wurthwein}},\ }in\ \href {\doibase
  10.1109/CSIE.2009.950} {\emph {\bibinfo {booktitle} {Computer Science and
  Information Engineering, 2009 WRI World Congress on}}},\ Vol.~\bibinfo
  {volume} {2}\ (\bibinfo {year} {2009})\ pp.\ \bibinfo {pages}
  {428--432}\BibitemShut {NoStop}%
\bibitem [{\citenamefont {Pordes}\ \emph {et~al.}(2007)\citenamefont {Pordes},
  \citenamefont {Petravick}, \citenamefont {Kramer}, \citenamefont {Olson},
  \citenamefont {Livny}, \citenamefont {Roy}, \citenamefont {Avery},
  \citenamefont {Blackburn}, \citenamefont {Wenaus}, \citenamefont
  {Würthwein}, \citenamefont {Foster}, \citenamefont {Gardner}, \citenamefont
  {Wilde}, \citenamefont {Blatecky}, \citenamefont {McGee},\ and\ \citenamefont
  {Quick}}]{Pordes2007}%
  \BibitemOpen
  \bibfield  {author} {\bibinfo {author} {\bibfnamefont {R.}~\bibnamefont
  {Pordes}}, \bibinfo {author} {\bibfnamefont {D.}~\bibnamefont {Petravick}},
  \bibinfo {author} {\bibfnamefont {B.}~\bibnamefont {Kramer}}, \bibinfo
  {author} {\bibfnamefont {D.}~\bibnamefont {Olson}}, \bibinfo {author}
  {\bibfnamefont {M.}~\bibnamefont {Livny}}, \bibinfo {author} {\bibfnamefont
  {A.}~\bibnamefont {Roy}}, \bibinfo {author} {\bibfnamefont {P.}~\bibnamefont
  {Avery}}, \bibinfo {author} {\bibfnamefont {K.}~\bibnamefont {Blackburn}},
  \bibinfo {author} {\bibfnamefont {T.}~\bibnamefont {Wenaus}}, \bibinfo
  {author} {\bibfnamefont {F.}~\bibnamefont {Würthwein}}, \bibinfo {author}
  {\bibfnamefont {I.}~\bibnamefont {Foster}}, \bibinfo {author} {\bibfnamefont
  {R.}~\bibnamefont {Gardner}}, \bibinfo {author} {\bibfnamefont
  {M.}~\bibnamefont {Wilde}}, \bibinfo {author} {\bibfnamefont
  {A.}~\bibnamefont {Blatecky}}, \bibinfo {author} {\bibfnamefont
  {J.}~\bibnamefont {McGee}}, \ and\ \bibinfo {author} {\bibfnamefont
  {R.}~\bibnamefont {Quick}},\ }\href
  {http://stacks.iop.org/1742-6596/78/i=1/a=012057} {\bibfield  {journal}
  {\bibinfo  {journal} {J. Phys: Conf. Ser.}\ }\textbf {\bibinfo {volume}
  {78}},\ \bibinfo {pages} {012057} (\bibinfo {year} {2007})}\BibitemShut
  {NoStop}%
\bibitem [{\citenamefont {Turner}\ \emph {et~al.}(2010)\citenamefont {Turner},
  \citenamefont {Zhang},\ and\ \citenamefont {Vishwanath}}]{Turner2010}%
  \BibitemOpen
  \bibfield  {author} {\bibinfo {author} {\bibfnamefont {A.~M.}\ \bibnamefont
  {Turner}}, \bibinfo {author} {\bibfnamefont {Y.}~\bibnamefont {Zhang}}, \
  and\ \bibinfo {author} {\bibfnamefont {A.}~\bibnamefont {Vishwanath}},\
  }\href {\doibase 10.1103/PhysRevB.82.241102} {\bibfield  {journal} {\bibinfo
  {journal} {Phys. Rev. B}\ }\textbf {\bibinfo {volume} {82}},\ \bibinfo
  {pages} {241102} (\bibinfo {year} {2010})}\BibitemShut {NoStop}%
\bibitem [{\citenamefont {Hughes}\ \emph {et~al.}(2011)\citenamefont {Hughes},
  \citenamefont {Prodan},\ and\ \citenamefont {Bernevig}}]{Hughes2011}%
  \BibitemOpen
  \bibfield  {author} {\bibinfo {author} {\bibfnamefont {T.~L.}\ \bibnamefont
  {Hughes}}, \bibinfo {author} {\bibfnamefont {E.}~\bibnamefont {Prodan}}, \
  and\ \bibinfo {author} {\bibfnamefont {B.~A.}\ \bibnamefont {Bernevig}},\
  }\href {\doibase 10.1103/PhysRevB.83.245132} {\bibfield  {journal} {\bibinfo
  {journal} {Phys. Rev. B}\ }\textbf {\bibinfo {volume} {83}},\ \bibinfo
  {pages} {245132} (\bibinfo {year} {2011})}\BibitemShut {NoStop}%
\bibitem [{\citenamefont {{Huang}}\ and\ \citenamefont
  {{Arovas}}(2012)}]{Huang2012}%
  \BibitemOpen
  \bibfield  {author} {\bibinfo {author} {\bibfnamefont {Z.}~\bibnamefont
  {{Huang}}}\ and\ \bibinfo {author} {\bibfnamefont {D.~P.}\ \bibnamefont
  {{Arovas}}},\ }\href@noop {} {\  (\bibinfo {year} {2012})},\ \Eprint
  {http://arxiv.org/abs/1205.6266} {arXiv:1205.6266} \BibitemShut {NoStop}%
\bibitem [{\citenamefont {Castelnovo}\ and\ \citenamefont
  {Chamon}(2007)}]{Castelnovo2007}%
  \BibitemOpen
  \bibfield  {author} {\bibinfo {author} {\bibfnamefont {C.}~\bibnamefont
  {Castelnovo}}\ and\ \bibinfo {author} {\bibfnamefont {C.}~\bibnamefont
  {Chamon}},\ }\href {\doibase 10.1103/PhysRevB.76.184442} {\bibfield
  {journal} {\bibinfo  {journal} {Phys. Rev. B}\ }\textbf {\bibinfo {volume}
  {76}},\ \bibinfo {pages} {184442} (\bibinfo {year} {2007})}\BibitemShut
  {NoStop}%
\end{thebibliography}%
\end{document}